\documentclass[preprint,5p]{elsarticle}

\usepackage{graphicx,subfig,bm}

\usepackage{amsmath}
\usepackage{amsthm}
\usepackage{eucal}
\usepackage{amssymb}
\usepackage{mathrsfs}
\usepackage{float}

 \newcommand{\eref}[1]{Eq.~(\ref{#1})}

\newcommand{\fref}[1]{figure \ref{#1}}

\newcommand{\Sref}[1]{Section \ref{#1}}

\newcommand{\sref}[1]{section \ref{#1}}

\begin{document}

\title{Population genetics on islands connected by an arbitrary network: An analytic approach}
\author{George W A Constable,  Alan J McKane}
\address{Theoretical Physics Division, School of Physics and Astronomy, The University of Manchester, Manchester M13 9PL, UK}
\begin{abstract}
We analyse a model consisting of a population of individuals which is subdivided into a finite set of demes, each of which has a fixed but differing number of individuals. The individuals can reproduce, die and migrate between the demes according to an arbitrary migration network. They are haploid, with two alleles present in the population; frequency independent selection is also incorporated, where the strength and direction of selection can vary from deme to deme. The system is formulated as an individual-based model, and the diffusion approximation systematically applied to express it as a set of nonlinear coupled stochastic differential equations. These can be made amenable to analysis through the elimination of fast-time variables. The resulting reduced model is analysed in a number of situations, including migration-selection balance leading to a polymorphic equilibrium of the two alleles, and an illustration of how the subdivision of the population can lead to non-trivial behaviour in the case where the network is a simple hub. The method we develop is systematic, may be applied to any network, and agrees well with the results of simulations in all cases studied and across a wide range of parameter values.
\end{abstract}

\begin{keyword}
Metapopulation \sep Migration \sep Selection \sep Moran model \sep Fast-mode reduction
\end{keyword}

\maketitle

\section{Introduction}\label{secIntroduction}

The founders of population genetics, who reconciled Mendelian genetics and Darwinian evolution, did so through the use of mathematical models which were frequently deterministic and which involved the key processes of mutation, migration and selection \cite{hartl1989}. Crucially, Fisher and Wright added genetic drift to this list by considering simple stochastic processes in systems where the population size, $N$, was finite \cite{fisher1930,wright1931}. Subsequent work tended to follow their original approach, assuming discrete generations and discrete state variables (corresponding to the number of individuals in the population carrying one particular type of allele)~\cite{ewens1969}. However adding more complexity to the models in the form of selection and migration makes this approach, based as it is on Markov chains, very unwieldy, and mathematical progress can be difficult~\cite{ewens2004}.

The solution to this dilemma is to take a mesoscopic perspective. That is, one uses a diffusion approximation in which the (discrete) number of individuals carrying a particular allele, $n$, is replaced by the (continuous) fraction $x=n/N$ carrying that allele \cite{crow2009}. Assuming $x$ is continuous is usually a good approximation for reasonably large $N$. This approximation, although originally suggested by Fisher \cite{fisher1922}, was popularised by Kimura \cite{crow1956,kimura1994}, and proved to be a powerful tool and the starting point for many studies of more complex processes in population genetics \cite{kimura1955,kimuraSSM,kimuraReview1964}. Nevertheless, some quite straightforward models give rise to rather complicated equations even within this approximation. For instance, a model with migration between $\mathcal{D}$ subpopulations (demes) which includes selection, leads to a nonlinear partial differential equation in $\mathcal{D}$ variables for the probability distribution function (pdf) \cite{mckaneModels2007} which seems quite intractable.

In this paper we show that equations such as these are in fact not as intractable as they seem, and in many cases can be reduced to a differential equation for a single variable, which can be straightforwardly analysed. The methodology which allows this reduction is based on the elimination of fast variables, and relies on a number of factors. First, rather than formulating the diffusion approximation in terms of a partial differential equation for the pdf, it is more useful to work in terms of an equivalent stochastic differential equation (SDE) \cite{mckaneSP}. This is a direct generalisation of the equation describing the deterministic dynamics \cite{mckaneBMB}. Second, this formulation of the system dynamics allows us to use much of the same intuition that is used to understand the deterministic process. In particular, we will see that in many of the cases of interest the dynamics can be divided into `fast' variables and `slow' variables \cite{serra1986}. After a short time, the dynamics of the fast variables may be ignored, since they have decayed to their stationary values; all the dynamics is contained in the few (in our case, frequently only one) slow mode. Third, the method is systematic and intuitive, and also applies to more general systems, such as those with many alleles and those involving other processes.

Our intention here is to apply this methodology to the case of a simple two-allele, haploid model with migration, even though it may be extended to deal with more complicated systems. Population genetics models featuring migration were first considered by Wright \cite{wright1931}, who looked at what is now often referred to as the standard island model \cite{rousset2004}. Instead of the well-mixed population of size $N$ which had been previously studied, he considered a set of $\mathcal{D}$ well-mixed subpopulations. With migratory individuals being chosen from the global population, there was no spatial structure assumed, only interactions between the various subpopulations. These are called demes in the genetic context \cite{hartl1989}, although within the modern nomenclature of ecology, these are effectively metapopulations \cite{levins1969}. Along with references to islands, we will use the terms demes and metapopulations interchangeably in this paper to refer to areas in which there is no spatial structure, but between which interactions can occur. The case of one deme therefore should reduce to the well mixed case.

Subsequently, the formulation of the stepping stone model \cite{kimuraSSM} introduced what was a very simple topology into the description of migration; the islands were ordered, with migration from island $i$ only allowed onto islands $i-1$ and $i+1$. Maruyama compared selection in the stepping stone model and in the island model \cite{maruyama1969}. He concluded that if selection was additive (i.e.~frequency independent) and local, with the same selection pressure in all demes, then deme population structure played no role. In other words, the population behaved approximately as a well-mixed and spatially homogeneous population, with a size equivalent to the sum of the deme population sizes.

The books by Ewens and Moran \cite{ewens2004,moran1962} describe variants of these models, analyses and conclusions, but for our purposes the next result of note is the work of Nagylaki \cite{nagylaki1980SM} who studied what would be in modern terminology an arbitrary network of demes. He constructed a migratory model with discrete generations (of the Wright-Fisher type) in the limit of strong migration, that is where the probability of a migration event is of the same order as that of a birth or death event.  The effect of this assumption was to create a separation of timescales in the Markov chain. Nagylaki then employed his earlier results on Markov chains with timescale separation \cite{nagylaki1980DIFF} to achieve an equation in the diffusion limit. Starting with a neutral model, it was concluded that in the long-time limit the population behaved as if it were well mixed, but with an effective population size less than or equal to the total unstructured population. Equality was shown to be achieved only if the migration matrix was symmetric.

The analysis which is used to reach these conclusions seems, to us at least, difficult to follow, with some parts of the proof relying on results from the theory of Markov chains and others relying in the nature of the diffusion approximation. Nevertheless, the results of the analysis are widely quoted and utilised. The work was extended \cite{nagylaki1980SM} to the case of different selection strengths on different islands, showing that once again the population was well-mixed with an effective population size, but now also with an effective selection coefficient. The situation where the selection on different islands operates in different directions was not discussed. In this case within certain parameter ranges a stable fixed point emerges, allowing coexistence of deleterious alleles in some demes, but disadvantageous in others. The deterministic implications of this have been discussed in \cite{moran1962,eyland1971,nagylaki2008}.

In the wake of this work a number of studies were carried out and a plethora of results obtained, all with a variety of different approximations and objectives. Several of these were concerned with an effort to determine the effective population size, which amounts to a rescaling of time for the structured population. Here we will avoid the temptation to describe the results that we obtain in terms of an effective population size, due to its amorphous definition, and at times misleading designation. We refer the reader to \cite{charlesworth2009} for a review of such work. Nagylaki's results on the diffusion limit of Markov chains with a separation of timescale have also been employed in \cite{sabin2009}, where they were used to good effect to extend the results to systems where fitness is not just additive, but frequency dependent. In this case, however, it becomes important to specify carefully between whom individuals compete (on their own island, on islands connected to their own, or the whole population) as these can sometimes lead to different results. In turn, other work has focused instead on the effect of migration on local deme properties \cite{blanquart2012}. 

The approach that we adopt in this paper will be to carefully define the model in terms of individuals (\textit{i.e.}~at the microscale). We will work within the context of continuous time Markov chains, that is, in terms of master equations \cite{gardiner2009}. We will therefore not assume non-overlapping generations as in the Wright-Fisher model, but instead work with the continuous time Moran process \cite{ewens2004}. As is well known, these two processes are essentially identical at medium to long times, up to a redefinition of time scales. The master equations for the Moran process involving birth, death, migration, mutation, and so on can be written down in a systematic way \cite{mckaneModels2007}, although it is too complicated to allow analytic progress to be made. As we have indicated the key to further progress is to write down a mesoscopic description which is achieved through a diffusion approximation which is derived by expanding the master equation in inverse deme size. 

A related set of questions to those that we ask here have been studied in a model of language evolution \cite{mckaneUtterance,mckaneConsensus,mckane2012}, in which each island is mapped on to a speaker having two different linguemes (different ways of saying the same thing) whose concentrations are modified through interaction events (analogous to migration events). While this model has similar features to the one we discuss here, it is distinct, and the methods of analysis and the final results are also different. We have already mentioned the work of Nagylaki \cite{nagylaki1980SM}. Once again our model, analysis and conclusions differ. Throughout this paper we will stress the systematic way that the underlying individual based model (IBM) can be constructed, and the straightforward and intuitive way that the mesoscopic version of this model can be reduced to an effective theory which can be analysed exactly. All approximations which we make will be checked through the use of numerical simulations in the form of the Gillespie algorithm~\cite{gillespie1976} applied to the underlying IBM.

In \Sref{secNeutralReview} we will introduce the model we use by studying the neutral Moran model on one island. We will illustrate the construction of the master equation and the way in which the mesoscopic description is obtained. We will also give the well-known results for the probability of fixation and mean time to fixation, which will be the quantities of interest in later sections. The generalisation to $\mathcal{D}$ demes will then be discussed in \Sref{secNeutralMigration}, and the corresponding mesoscopic description obtained. Similarly in \Sref{secSelectionReview} the one island Moran model with selection will be reviewed, before the general model with migration and selection is described in \Sref{secSelectionMigration}.

In \Sref{secReduction}, we give a description of the model reduction method in both the neutral case, and the case with selection. The mathematical derivation of the results we use is given in \cite{projectionPhys}. Here we simply state the one-dimensional reduced equation that approximates the full system. In \Sref{secApplications} we explore the predictions of the reduced model, calculating the probability of fixation and the mean time to fixation in the neutral case in \Sref{secProbabilitiesNeutral} and the case with selection in \Sref{secProbabilitiesSelection}. We find the approximation captures the behaviour of the system remarkably well. We then proceed to apply these results on the probability and time to fixation to two specific systems of interest. 

The first, discussed in \Sref{secMigrationSeletionBalance}, is a system in which migration and selection balance to induce a polymorphic equilibrium. To our knowledge such results have only been obtained previously for standard island models \cite{tachida} and two deme cases with symmetric migration \cite{gavrilets2002}, both of which form a subset of of the cases we address here. While work by Whitlock \cite{whitlock2005} allows asymmetric migration and multiple demes, the selection strength may only take on two distinct values in those many demes.

Finally, in \Sref{secHub}, we will give an illustrative example of the predictive power of the reduction by considering the  case of a `hub' or `spoke' topology, where a central island is connected to $(\mathcal{D}-1)$ other islands, none of which are connected to each other. The subdivision of the population is seen to have a non-trivial effect on the system which the reduced model accurately predicts.

Three appendices are given in which technical details are discussed. The first, \ref{appKMExpansion}, covers the Kramers-Moyal expansion of the microscopic model, the second, \ref{appSol}, the general calculation of the probability of fixation and the third, \ref{appMTF}, the calculation of the mean time to fixation.

\section{The migration models}\label{secModels}

\subsection{The neutral Moran model and the diffusion limit}\label{secNeutralReview}

In order to make our analysis of the Moran model with coupled migration clear, we first briefly review the well-mixed Moran model. We begin by formulating this as an IBM with a dynamics given by a master equation, before moving to the continuous limit, in which we arrive at a Fokker-Planck equation (FPE). Readers who wish to see mathematical details of these stages can consult \ref{appKMExpansion}. In the main text however we will attempt to restrict attention to the key results and conceptual ideas behind the techniques we will use. Throughout we will make explicit any assumptions made, or relationships inferred in the simplification and analysis of the system.

The population we consider is finite, well-mixed and composed of haploid individuals containing one of two alleles, $A$ and $B$; the number of each is given by the integers $n$ and $m$ respectively. At a specific point in time we pick an individual to reproduce. We assume that the progeny of this reproduction event carries the allele of its parent and that it immediately displaces a pre-existing individual at random. In this way the population size is kept constant so that at any one time $n+m=N$, where $N$ is the total size of the well-mixed population. This is the simplest version of the Moran model~\cite{moran1957,moran1962}. A set of transition rates can then be defined which describe the probability per unit time that allele $A$ increases or decreases in the population. In this case the transition rates may be obtained from simple combinatorics, and are given by 
\begin{eqnarray*}
 T(n+1|n) = \frac{1}{N(N-1)}n(N-n) \,,\\
 T(n-1|n) = \frac{1}{N(N-1)}n(N-n) \,.
\end{eqnarray*}
These, together with the master equation~\cite{vankampen2007}
 \begin{eqnarray}\label{masterEquation1D}
 \frac{d p(n,t)}{d t} &=& T(n|n-1)p(n-1,t) \nonumber \\ &+& T(n|n+1)p(n+1,t) \nonumber \\ &-& \left[T(n+1|n) + T(n-1|n) \right] p(n,t) \,, 
\end{eqnarray}
define the evolution in time of the probability distribution $p(n,t)$. The interpretation of the master equation is intuitively clear: the probability that the system is in state $n$ increases with the probability that the system moves into it from one of the surrounding states, $n-1$ or $n+1$,  but decreases with the probability that the system is already in state $n$ but transitions to another state. Despite this simple description, the master equation is very rarely analytically tractable. We must resort to solving instead an approximation of the equation. 

The diffusion approximation leads to a more tractable equation than the full master equation. It involves an expansion in inverse system size, and is valid in the limit of large $N$. First we change variables to $x=n/N$ and Taylor expand the governing master equation in powers of $N^{-1}$. Formally this is known as the Kramers-Moyal expansion~\cite{gardiner2009,risken1989}. Truncating the series at second order in $N^{-1}$ leads to the FPE which has the form~\cite{mckaneBMB}
\begin{eqnarray}\label{FPE1D}
 \frac{\partial p(x,t)}{\partial t} = &-& \frac{1}{N}\frac{\partial}{\partial x} \left[A(x)p(x,t)\right] \nonumber \\ &+& \frac{1}{2N^{2}}\frac{\partial^{2}}{\partial x^{2}} \left[B(x)p(x,t)\right] \,, 
\end{eqnarray}
where the functions $A(x)$ and  $B(x)$ are given respectively by
\begin{eqnarray}\label{1Ddiffusion}
 A(x) = 0 \, , \quad \, B(x) = 2 x(1-x) \,.
\end{eqnarray}
It is common practice~\cite{moran1962,ewens2004} to rescale time by introducing $\tau$ such that $t = N \tau $, but here we retain the $N$ dependence for clarity. 

\begin{figure}[h!]
\centering
\includegraphics[width=0.23\textwidth]{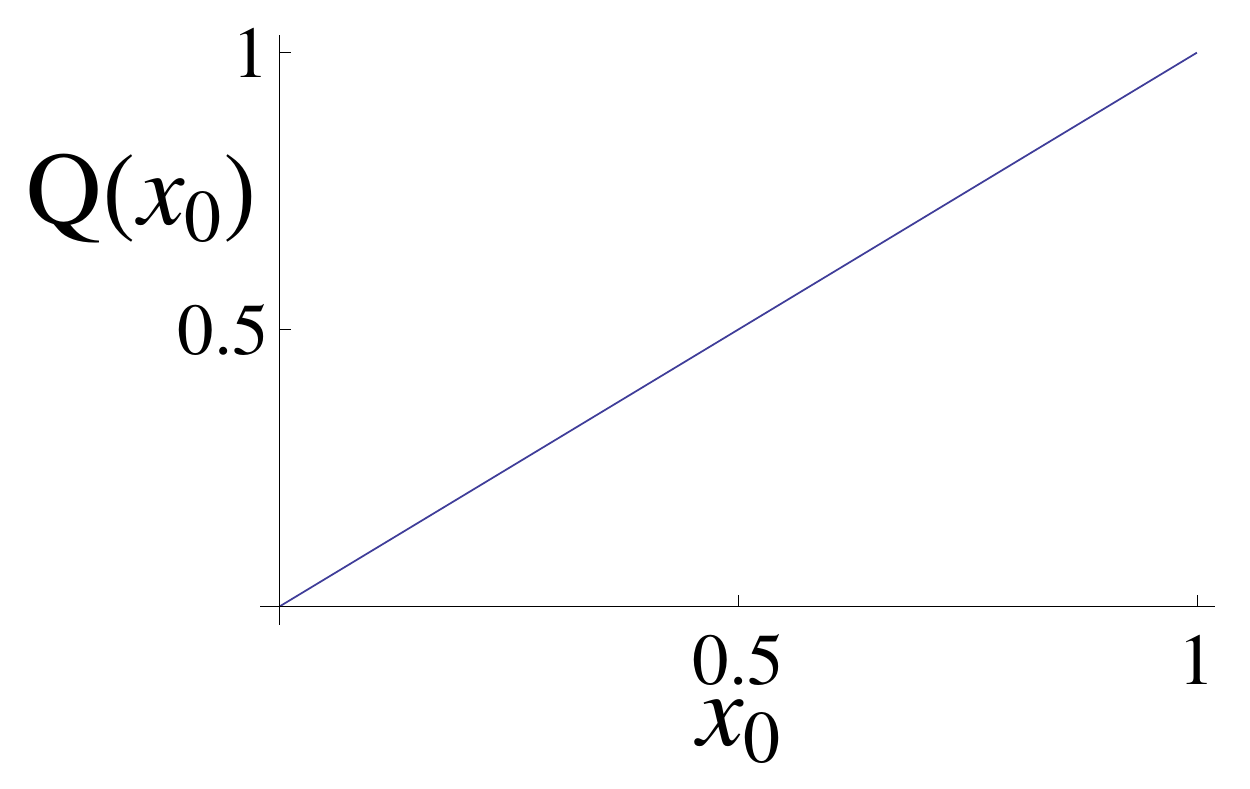}
\includegraphics[width=0.23\textwidth]{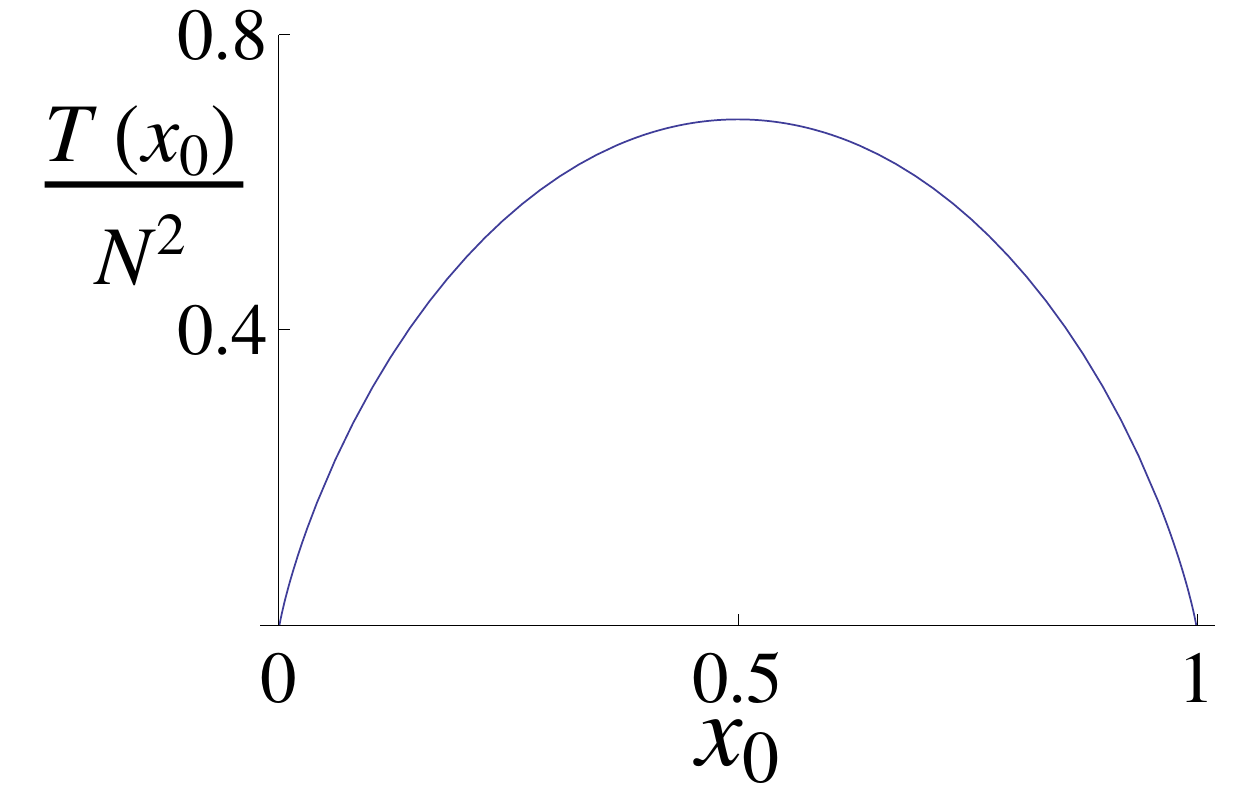}

\caption{Graphical summary of results from a well-mixed Moran model without selection. Left panel; probability of fixation of allele $A$, $Q(x_{0})$, in a neutral system as a function of initial $A$ allele concentration, $x_{0}$. Right panel; time to fixation of either $A$ or $B$ allele, $T(x_{0})$, scaled by the system size squared, $N^{2}$, in a neutral system as a function of the initial concentration of $A$ allele.}
\label{1DNeutralSummary}
\end{figure}

A more useful starting point for calculating many quantities of interest is not the FPE itself, but the backward Fokker-Planck equation, which is formally the adjoint of the FPE~\cite{gardiner2009,risken1989}. In the usual, or forward, FPE we impose initial conditions at time $t_0$  and find $p(x,t)$ for $t>t_0$. In the backward FPE we impose conditions at time $t$ and ask what solution $q(x_0,t_0)$ at an initial time $t_0 <t$ gave this final condition. For this reason the backward equation is most useful in the investigation of, for instance, the probability of the fixation of allele $A$ given some initial concentration $x_{0}$, denoted by $Q(x_{0})$, and the mean time it takes either allele $A$ or $B$ to fixate, $T(x_{0})$. Ordinary differential equations for $Q(x_0)$ and $T(x_0)$ can be found from the backward FPE. Details are given in many standard texts: general diffusion problems are described in~\cite{gardiner2009,risken1989}, whilst the methods for the specific case under consideration here are illustrated in~\cite{ewens1969}. After some calculation~\cite{moran1962,ewens2004,gardiner2009}, one finds the equations 
\begin{eqnarray}\label{BFPQ}
\frac{A(x_0)}{N}\,\frac{dQ(x_{0})}{dx_{0}} + \frac{B(x_{0})}{2N^2}\,\frac{d^{2}Q(x_{0})}{dx_{0}^{2}} = 0 \,, \\
\end{eqnarray}
with boundary conditions $Q(0)=0$, $Q(1)=1$, and 
\begin{eqnarray}\label{BFPT}
\frac{A(x_0)}{N}\,\frac{dT(x_{0})}{dx_{0}} + \frac{B(x_{0})}{2N^2}\,\frac{d^{2}T(x_{0})}{dx_{0}^{2}} = -1 \,,
\end{eqnarray}
with boundary conditions $T(0)=0$, $T(1)=0$.
For this neutral case the solutions are~\cite{ewens2004}
\begin{eqnarray}
 Q(x_0) &=& x_0 \,, \label{1D_Q_s_0} \\
 T(x_{0})&=& - N^2 \left[  (1-x_{0})\ln{(1 - x_{0})}  +  x_0 \ln{(x_0)} \right] \, . \nonumber \\ \label{1D_T_s_0}
\end{eqnarray}
These results are shown graphically in \fref{1DNeutralSummary}.


\subsection{Definition of the neutral metapopulation Moran model}\label{secNeutralMigration}

This model consists of a series of $\mathcal{D}$ demes, on each of which well-mixed populations of fixed and finite size exist. The number of individuals island $i$ contains is given by $\beta_{i}N$, where $N$ is some typical island size, and $\beta_{i}$ is a scaling factor such that $\beta_{i}N$ is an integer. The individuals in the population can carry one of two alleles, $A$ and $B$. An independent deme, unconnected to any others and with a sufficiently large population size, would then be well described by the FPE~(\ref{FPE1D}). However we are interested in the form of the FPE for the system which comprises the whole set of $\mathcal{D}$ demes, with migration between them. 

\begin{figure}[h!]

\fbox{ \subfloat[Subfigure 1 list of figures text][]{
\includegraphics[width=0.21\textwidth]{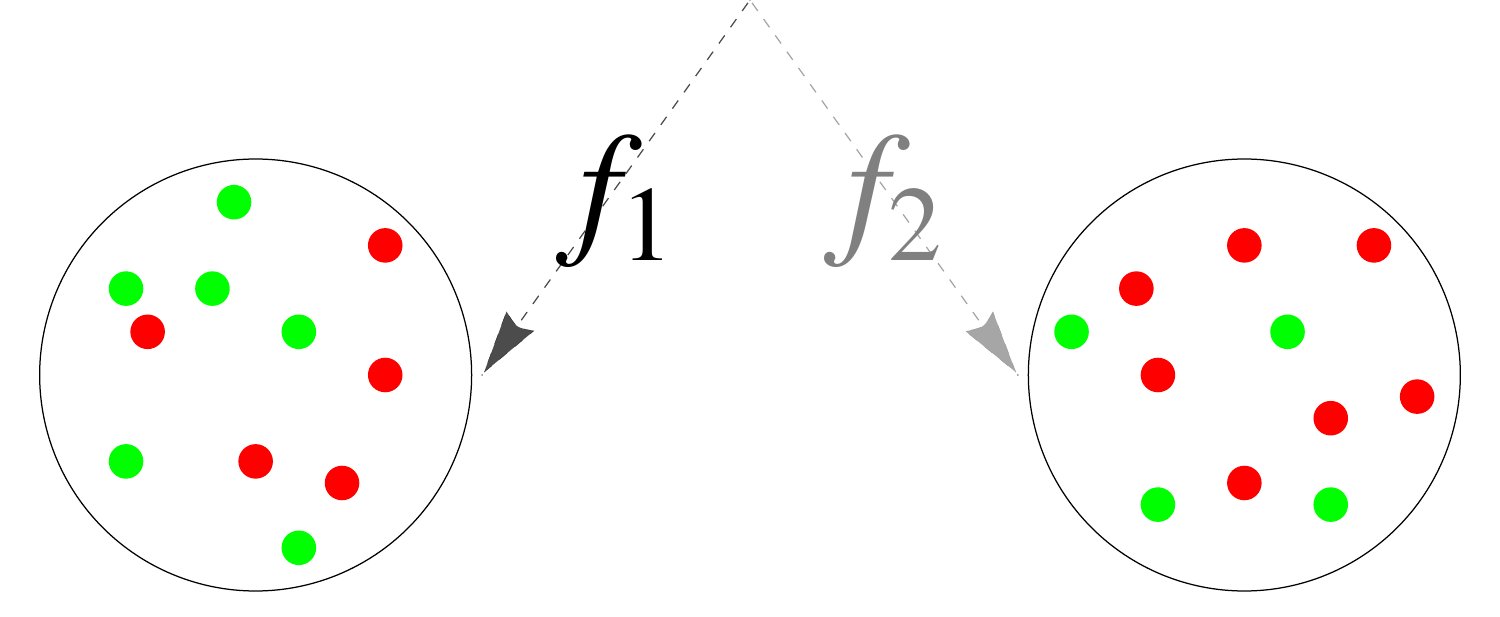}
\label{fig:subfig1}} }   \fbox{ \subfloat[Subfigure 2 list of figures text][]{\includegraphics[width=0.21\textwidth]{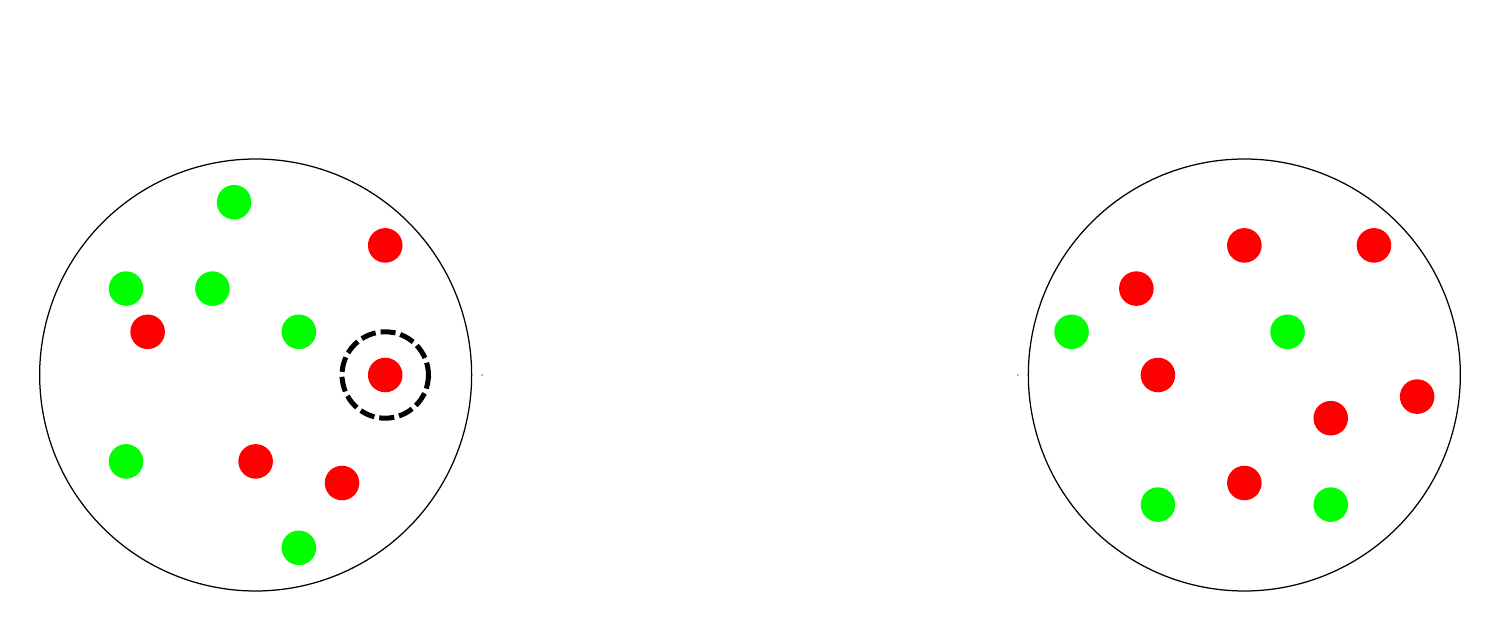}\label{fig:subfig2}} }

\fbox{ \subfloat[Subfigure 3 list of figures text][]{
\includegraphics[width=0.21\textwidth]{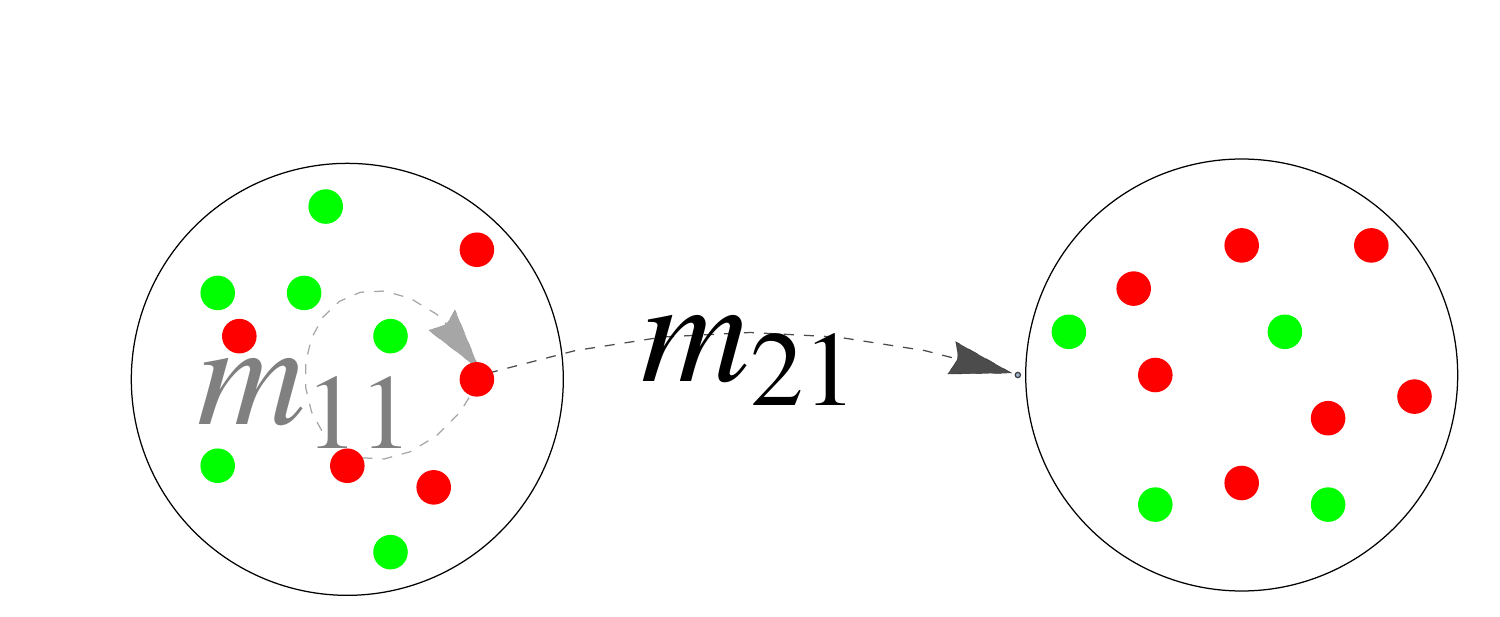}
\label{fig:subfig3}} }   \fbox{ \subfloat[Subfigure 4 list of figures text][]{\includegraphics[width=0.21\textwidth]{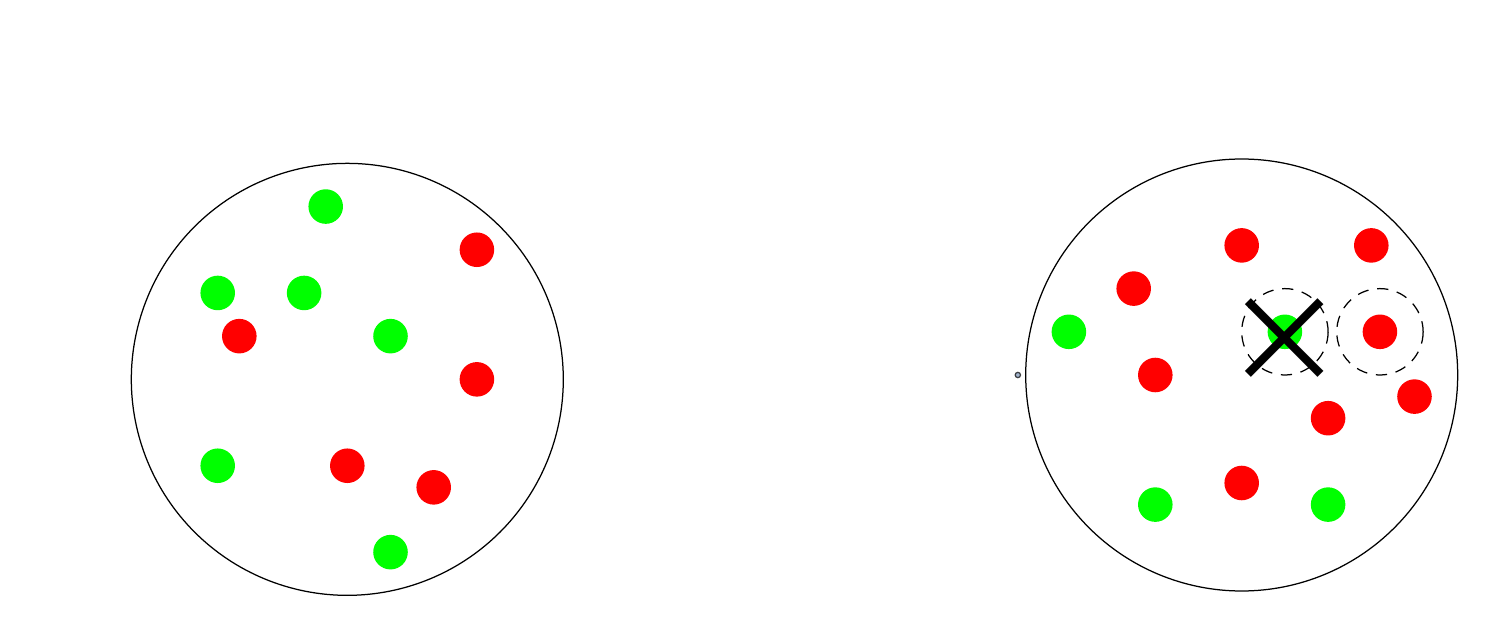}\label{fig:subfig4}} }

\caption{ Diagram of the neutral metapopulation Moran model for $\mathcal{D}=2$. Each large black circle is a deme populated by two types of haploid individuals carrying either an allele $A$, in red, or allele $B$ in green. Subfigures (a)-(d) depict the stages in picking an allele to reproduce and an allele to die.}
\label{neutralDiagram}
\end{figure}

The process is shown diagrammatically in \fref{neutralDiagram} for the case $\mathcal{D}=2$. In \fref{neutralDiagram}(a), a reproduction site is chosen with probability $f_{j}$, which corresponds to a total birth rate for deme $j$; if the demes have an equal birth rate per captia, we simply find $f_{j}=\beta_{j}(\sum_{i=1}^{\mathcal{D}} \beta_{i})^{-1}$. In \fref{neutralDiagram}(b) either one of the two alleles is chosen to reproduce based on their relative frequencies in that deme. The individual then reproduces and its progeny may either displace an individual in their own deme, or an individual in another deme according to the matrix element $m_{ij}$ (see \fref{neutralDiagram}(c)). The matrix $m_{ij}$ is then the probability that a individual reproducing in $j$ will have offspring which displaces an individual in $i$. Finally, in \fref{neutralDiagram}(d) the type of individual in $i$ that is displaced is decided, again based on their relative frequencies in $i$. The vector $f_{j}$ and the matrix $m_{ij}$ represent probabilities and so satisfy the conditions $\sum_{j}f_{j}=1$ and $\sum_{i}m_{ij}=1$ for all $j$.

 
We denote the number of $A$ alleles in deme $i$ by $n_i$. The number of $B$ alleles in deme $i$ is then given by $\beta_{i}N-n_{i}$, where $\beta_{i}N$ is the total population of that deme. The equivalent transition rates for the process depicted in \fref{neutralDiagram} can then be calculated using combinatoric arguments. Starting with deme $i$ and summing over the $\mathcal{D}$ demes from which an allele can originate, one obtains
\begin{eqnarray}
 T(n_{i}+1|n_{i}) &=& \sum_{j=1}^{\mathcal{D}} \left( f_{j} \right) \left(\frac{n_{j}}{\beta_{j} N} \right) \times \nonumber \\ && \left( m_{ij} \right) \left(\frac{\beta_i N - n_{i}}{\beta_{i}N - \delta_{ij}} \right) \, , \nonumber \\ 			
   T(n_{i}-1 |n_{i}) &=& \sum_{j=1}^{\mathcal{D}} \left( f_{j} \right) \left(\frac{\beta_{j}N-n_{j}}{\beta_{j}N} \right) \times \nonumber \\ &&  \left( m_{ij} \right) \left(\frac{n_{i}}{\beta_{i}N - \delta_{ij}} \right) \, , \nonumber 						
\end{eqnarray}
where the dependence of $T(\bm{n}|\bm{n}')$ on the elements of $\bm{n}$ that do not change in the transition have been suppressed. Each of the four factors in these expressions for $T(n_{i}\pm1|n_{i})$ corresponds to one of the four processes displayed in \fref{neutralDiagram}. The diagonal elements of $m_{ij}$ do not represent a migration process, but instead the probability that an offspring remains in its parent's deme; they are simply equal to one minus the sum of the other elements in the same column to ensure that $\sum_i m_{ij}=1$. 

Since $f_j$ and $m_{ij}$ always occur together in the combination $m_{ij}f_{j}$, it is convenient to introduce the matrix $G_{ij} \equiv m_{ij}f_{j}$, which we shall call the migration rate matrix, a combination of birth and migratory rates. The migration rate matrix inherits the properties $\sum_{i}G_{ij}=f_{j}$ and $\sum_{i,j}G_{ij}=1$. Again, the diagonal elements of $G_{ij}$ represent the probability that both deme $j$ is chosen \emph{and} the progeny of the reproduction remains in the parent deme. 

The transition rates in terms of $G_{ij}$ then become
\begin{eqnarray}
T(n_{i}+1|n_{i}) &=&   G_{ii} \frac{( \beta_{i} N - n_{i})n_{i}}{(\beta_{i}N-1)\beta_{i}N}  \nonumber \\ &+& \frac{\beta_i N - n_{i}}{\beta_{i}N} \sum_{j\neq i}^{\mathcal{D}} G_{ij}\frac{n_{j}}{\beta_{j}N} \,, \label{transitionRatesNeutral1} \\
T(n_{i}-1 | n_{i}) &=&  G_{ii} \frac{(n_{i} )(\beta_{i}N-n_{i})}{(\beta_{i}N - 1)\beta_{i}N} \nonumber \\ &+& \frac{n_{i}}{\beta_{i}N} \sum_{j\neq i}^{\mathcal{D}} G_{ij} \frac{\beta_{j}N-n_{j}}{\beta_{j}N} \, , \label{transitionRatesNeutral2}
\end{eqnarray}
where we have separated out the contribution from the processes involving two islands ($i$ and $j$) from those which only involve island $i$.  

The master equation associated with this process \newline (\eref{Meqn} given in \ref{appKMExpansion}) is clearly more complicated that that for the well-mixed single-island population, \eref{masterEquation1D}, and no more tractable. Again however, we can make the diffusion approximation. This time we make the change of variables $x_{i}=n_{i}/(\beta_{i}N)$ and again expand in powers of $N^{-1}$ (see \ref{appKMExpansion}, with the parameter $s$ set to zero). Truncating at second order we have the FPE for the metapopulation:
\begin{eqnarray}\label{generalFPE}
 \frac{\partial p(\bm{x},t)}{\partial t} = &-& \frac{1}{N} \sum_{i=1}^{\mathcal{D}} \frac{\partial}{\partial x_{i}} \left[A_{i}(\bm{x})p(\bm{x},t)\right] \nonumber \\ &+& \frac{1}{2N^{2}}\sum_{i=1}^{\mathcal{D}}\frac{\partial^{2}}{\partial x_{i}^{2}} \left[B_{ii}(\bm{x})p(\bm{x},t)\right],
\end{eqnarray}
with 
\begin{eqnarray}\label{neutralDrift}
 A_{i}(\bm{x}) =  \frac{1}{\beta_{i}} \left(- x_{i}\sum_{j \neq i}^{\mathcal{D}}G_{ij} + \sum_{j \neq i}^{\mathcal{D}}G_{ij}x_{j} \right)\, ,
\end{eqnarray}
and
\begin{eqnarray}\label{neutralDiff}
 B_{ii}(\bm{x}) &=&   \frac{1}{\beta_{i}^{2}}\left(x_{i}\sum_{j=1}^{\mathcal{D}}G_{ij} + \sum_{j=1}^{\mathcal{D}}G_{ij}x_{j} \right. \nonumber \\ & & \hspace{2cm} \left. - 2x_{i}\sum_{j=1}^{\mathcal{D}}G_{ij}x_{j} \right) \,.
\end{eqnarray}

It is sometimes assumed that the off-diagonal elements of the matrix $G_{ij}$ are such that $G_{ij} = \mathcal{G}_{ij}/N$ for all $i \neq j$, where $\mathcal{G}$ is of order unity \cite{nagylaki1980SM,mckaneModels2007}. This means that the off-diagonal elements in $B(\bm{x})$ may be neglected, since they give $\mathcal{O}(N^{-3})$ contributions. Since only the off-diagonal elements of $G$ appear in the vector $\bm{A}(\bm{x})$ and only the diagonal elements of $G$ appear in $B(\bm{x})$, both terms on the right-hand side of Eq.~(\ref{generalFPE}) are of order $N^{-2}$, and they effectively balance each other. Asking that the off-diagonal elements of the matrix $G$ are small has a clear biological interpretation. The population is strongly subdivided and it is far more likely that an individual's offspring will remain in the deme of its parents than migrate. This is not the generic case however, and the scaling of the off-diagonal terms with the inverse of the population size is in some cases little more than a mathematical convenience.

Here we will make the choice that the elements of the migration matrix are approximately all of the same order. In doing so we are assuming that once a deme is selected, the probability of allele reproduction-migration is not too dissimilar to that of allele reproduction. 

\subsection{The Moran model with selection}\label{secSelectionReview}

To add further complexity to the model prescribed by \Sref{secNeutralReview} one may add the effect of fitness. Specifically we shall incorporate frequency independent fitness, that is, fitness that does not depend on the constitution of the population. The variables $w_{A}$ and $w_{B}$ are introduced to represent the fitness weightings of alleles $A$ and $B$ respectively \cite{nowak2006}. The transition rates (see \ref{appKMExpansion}) are then 
\begin{eqnarray}\label{1DSelectionTransitions}
 T(n+1|n) &=& \frac{n w_A}{n w_A + (N-n)w_B}\frac{(N-n)}{N} \, ,\\
 T(n-1|n) &=& \frac{(N-n) w_B}{n w_A + (N-n)w_B}\frac{n}{N} \, . 
\end{eqnarray}
The appearance of $n$ in the denominator complicates the Kramers-Moyal expansion slightly. This is usually addressed by rewriting the fitness parameters $w_{A}=1+s$ and $w_{B}=1$, and expanding in powers of $s$ under the very reasonable assumption that $s$ is small. Positive $s$ indicates a small fitness advantage for individuals with allele $A$. 

The diffusion approximation leads to the FPE~(\ref{FPE1D}), with $A(x)$ and $B(x)$ given respectively by 
\begin{eqnarray}\label{DriftDiffSelection1D}
A(x) = sx(1-x) \quad \mathrm{and} \quad B(x) = 2 x(1-x) \,,
\end{eqnarray}
for small $s$. Here we have included only the lowest order contribution in $s$ to $A(x)$, and omitted the order $s$ correction in $B$ altogether, since it will be negligible compared with $2x(1-x)$ (for further details see \ref{appKMExpansion}). In the same manner as in \Sref{secNeutralReview} we find equations \eref{BFPQ} and \eref{BFPT}, with $A(x)$ and $B(x)$ given by \eref{DriftDiffSelection1D}, for the fixation probability $Q(x_{0})$ and the fixation time $T(x_{0})$ as a function of $x_0$, the initial concentration of allele $A$.

\begin{figure}[h!]
\centering
\includegraphics[width=0.22\textwidth]{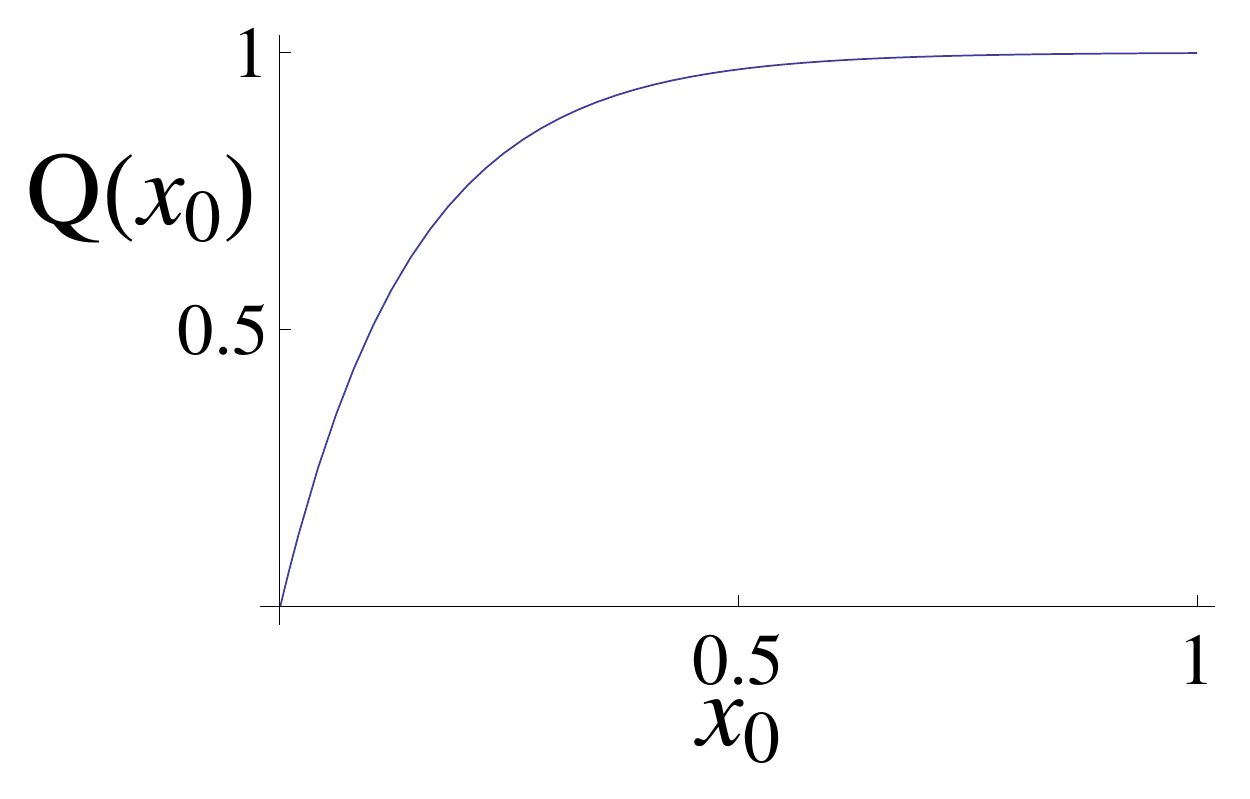}
\includegraphics[width=0.22\textwidth]{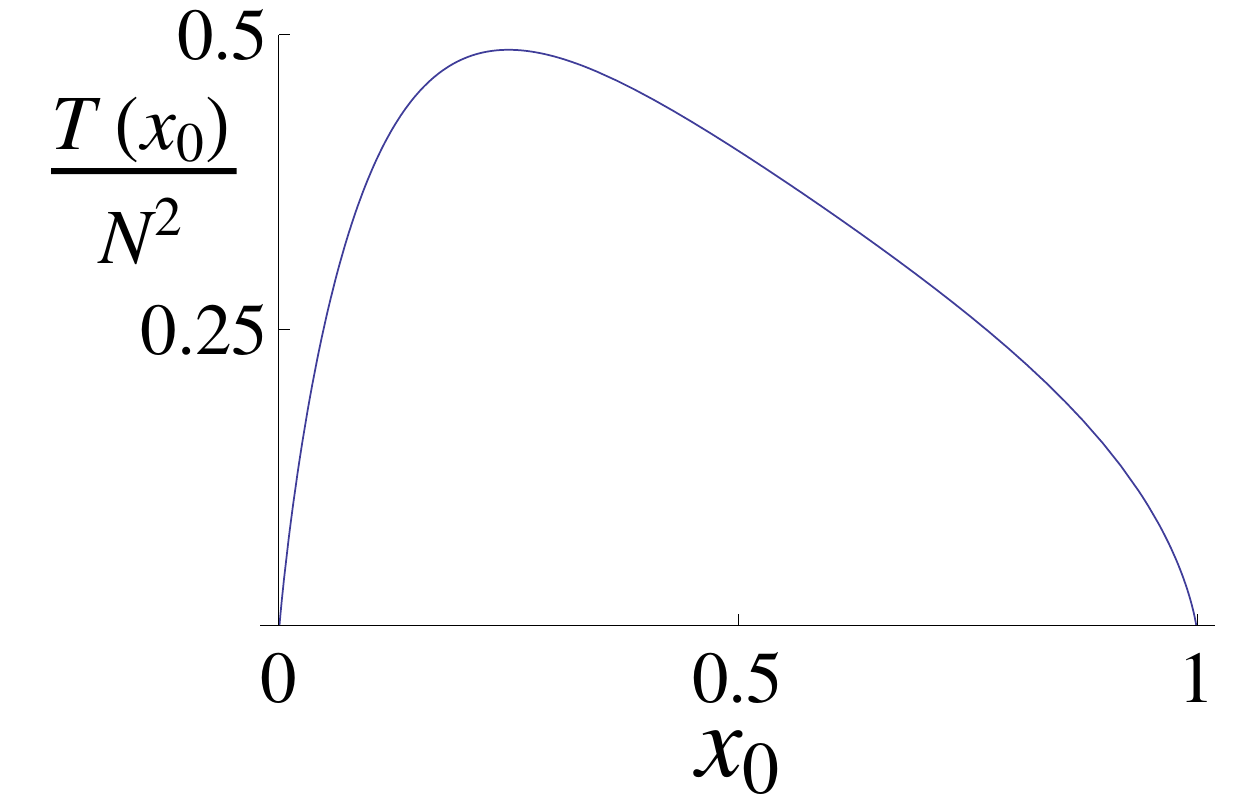}
\caption{Probability of fixation of allele $A$, $Q(x_{0})$, and mean time to fixation of either allele, $T(x_{0})$, in a system where allele $A$ has a selective advantage $s$ over allele $B$, as a function of initial $A$ concentration. While $Q(x_{0})$ is obtained from \eref{1D_Q_s_N}, $T(x_{0})$ has been obtained numerically.}
\label{1DFitnessSummary}
\end{figure}

In this case, solving \eref{BFPQ}, the familiar equation for the probability of fixation, $Q(x_{0})$ is found~\cite{ewens2004}:
\begin{eqnarray}\label{1D_Q_s_N}
 Q(x_{0}) = \frac{1 - \exp{(- N s x_{0})}}{1 - \exp{(- N s)}}.
\end{eqnarray}
While the mean time to fixation can also be obtained analytically (see Eqs.~(\ref{T_linear_in_s}) and (\ref{c_2}) with $M$ set to $N$ and $\sigma$ set to $s$,
in Section \ref{secProbabilitiesSelection}), it is sufficient to determine it numerically. Illustrative plots are shown in \fref{1DFitnessSummary}.


\subsection{The Moran model with fitness and migration}\label{secSelectionMigration}

We now proceed to incorporate selection into the migration model defined in \Sref{secNeutralMigration}. We begin as before with a set of transition rates, however now  the probability of an individual coming into deme $i$ from deme $j$ is a function of the progenitor's fitness in deme $j$. In a similar fashion to the case presented in \Sref{secSelectionReview}, the fitness of allele $A$ on each deme is denoted by the vector $\bm{w}_{A}$ while the fitness of allele $B$ is $\bm{w}_{B}$. Further details are given in \ref{appKMExpansion}, though the progression from the one deme model to the $\mathcal{D}$ deme model is straightforward. The transition rates are
\begin{eqnarray}\label{transitionRatesSelection}
 & & \hspace{-1cm} T( n_{i}+1| n_{i} ) = \nonumber \\  && \hspace{-0.5cm} \sum_{j=1}^{\mathcal{D}}  \frac{(\beta_i N-n_{i})}{\beta_{i}N - \delta_{ij}}G_{ij} \frac{ [\bm{w}_{A}]_{j}n_{j}}{[\bm{w}_{A}]_{j}n_{j}+  [\bm{w}_{B}]_{j}(\beta_{j}N-n_{j})}, \nonumber \\
 & & \hspace{-1cm}T(n_{i}-1 |  n_{i} ) = \nonumber \\  && \hspace{-0.5cm}\sum_{j=1}^{\mathcal{D}}   \frac{n_{i}}{\beta_{i}N-\delta_{ij}}G_{ij}\frac{ [\bm{w}_{B}]_{j}(\beta_{j}N- n_{j})}{[\bm{w}_{A}]_{j}n_{j}+  [\bm{w}_{B}]_{j}(\beta_{j}N-n_{j})} \,.
\end{eqnarray}
Letting $[\bm{w}_{B}]_{i}=1$ for every island $i$, the elements of the fitness term $[\bm{w}_{A}]$ are now dependent on both the typical selection strength, $s$, and the vector $\bm{\alpha}$, which moderates the typical selection strength in magnitude and direction such that
\begin{equation}
 [\bm{w}_{A}]_{i} = 1 + s \alpha_{i}.
\end{equation}
Positive $\alpha_{i}$ therefore corresponds to allele $A$ being advantageous relative to $B$ on island $i$, while negative $\alpha_{i}$ means $A$ is deleterious. We assume that the elements of $\bm{\alpha}$ are of order unity.

Proceeding in the in the same spirit as \Sref{secNeutralMigration}, we conduct a Kramers-Moyal expansion to arrive at Fokker-Planck equation \eref{generalFPE} valid in the limit of large $N$ and small $s$. The calculation is carried out in full in \ref{appKMExpansion}. The FPE is defined through an $\bm{A}(\bm{x})$ vector and a diagonal $B(\bm{x})$ matrix which, when expressed as a series in $s$, have elements
\begin{flalign}\label{generalSelectionDrift}
 A_{i}(\bm{x}) &= \frac{1}{\beta_{i}}\left \{ \vphantom{\sum_{j=1}^{\mathcal{D}} G_{ij}\alpha_{j}^{2}x_{j}^{2}(1-x_{j}) } \sum_{j\neq i}^{\mathcal{D}} G_{ij}(x_{j} - x_{i}) + s \sum_{j=1}^{\mathcal{D}} G_{ij}\alpha_{j}x_{j}(1-x_{j})  \right. \nonumber \\
                       &\left. - s^{2} \sum_{j=1}^{\mathcal{D}} G_{ij}\alpha_{j}^{2}x_{j}^{2}(1-x_{j})  \right \} + \mathcal{O}(s^{3}) \, .
\end{flalign}
and 
\begin{eqnarray}\label{generalSelectionDiff}
 B_{ii}(\bm{x}) &=& \frac{1}{\beta_{i}^{2}} \left\{  x_{i}\sum_{j=1}^{\mathcal{D}}  G_{ij}  +\sum_{j=1}^{\mathcal{D}} G_{ij}x_{j}  \right. \nonumber \\ & & \hspace{1.cm} \left. - 2 x_{i}\sum_{j=1}^{\mathcal{D}}  G_{ij}  x_{j}         \right\} +  \mathcal{O}(s) \, .
\end{eqnarray}

\section{The approximation procedure }\label{secReduction}

Having derived the FPEs for a sequence of progressively more complex situations, we can now describe the approximation which is the main subject of this paper --- the reduction of the full model to an effective one-dimensional system. The full details of the approximation method and calculation are given in \cite{projectionPhys}. Here we will give a description of the technique before stating the equation of the reduced system, and then applying it.

The reduction technique relies on utilising concepts from deterministic dynamical systems theory in order to understand and simplify the stochastic equations. For this reason it is more convenient to work in the context of SDEs, rather than the FPE. In general, the system described by the FPE~(\ref{generalFPE}), with $\bm{A}(\bm{x})$ given by \eref{generalSelectionDrift} and $B(\bm{x})$ by \eref{generalSelectionDiff}, is entirely equivalent to the It\={o} SDE~\cite{gardiner2009,risken1989}
\begin{equation}\label{generalSDE}
\dot{x}_{i} = A_{i}(\bm{x}) + \frac{1}{\sqrt{N}}\eta_{i}(\tau) \, ,
\end{equation}
where the dot indicates differentiation with respect to $\tau=t/N$ and $\bm{\eta}(\tau)$ is a Gaussian white noise with zero mean and correlation functions
\begin{equation}
\langle \eta_{i}(\tau)\eta_{j}(\tau') \rangle = B_{ij}(\bm{x}) \delta(\tau-\tau') \, . 
\label{eta_correlator}
\end{equation}
We may intuitively think of this as the deterministic system, $\dot{x}_{i} = A_{i}(\bm{x})$, with a small amount of added noise. Before proceeding it is useful to re-write $A_{i}(\bm{x})$, highlighting its linearity at zeroth order in $s$:
\begin{eqnarray}
 \hspace{-0.5cm} A_{i}(\bm{x}) = && \hspace{-0.5cm} \sum_{j=1}^{\mathcal{D}}\,H_{ij}x_{j}  + \beta^{-1}_{i}\left \{ s\vphantom{\sum_{j=1}^{\mathcal{D}} G_{ij}\alpha_{j}^{2}x_{j}^{2}(1-x_{j}) } 
							      \sum_{j=1}^{\mathcal{D}} G_{ij}\alpha_{j}x_{j}(1-x_{j})  \right. \nonumber \\ 
                       && \hspace{1.5cm} \left. -s^{2}\sum_{j=1}^{\mathcal{D}} G_{ij}\alpha_{j}^{2}x_{j}^{2}(1-x_{j})  \right \} \, ,
\label{A_second_order}
\end{eqnarray} 
where the matrix $H$ has elements
\begin{equation}\label{defineH}
 H_{ij}=\frac{G_{ij}}{\beta_{i}} \ \ \mathrm{if\ } i \neq j, \qquad H_{ii} = -\sum_{j \neq i}^{\mathcal{D}} \frac{G_{ij}}{\beta_{i}} \, .
\end{equation}
The neutral system is then clearly obtained by setting $s=0$.

The properties of the matrix $H$ are central to the application of our method, and so we briefly summarise them. In general $H$ will not be symmetric, and so its eigenvalues may be complex, and it will have distinct right- and left-eigenvectors. We will denote the right- and left-eigenvectors corresponding to the eigenvalue $\lambda^{(i)}$ as $\bm{v}^{(i)}$ and $\bm{u}^{(i)}$ respectively. They are orthogonal if they correspond to different eigenvalues and can be defined so that they are orthonormal:
\begin{equation}
\sum_{k=1}^{\mathcal{D}}v^{(i)}_{k}u^{(j)}_{k} = \delta_{ij}\,.
\label{orthonormal}
\end{equation}
Furthermore, the first eigenvalue of $H$, $\lambda^{(1)}$, is zero, and the first right eigenvector has components $v^{(1)}_{i}=1$ for all $i$ and for any choice of parameters. All other eigenvalue's of $H$ can be shown to have a negative real part~\cite{projectionPhys}.

\begin{figure}[h!]
\centering
 \includegraphics[width=0.4\textwidth]{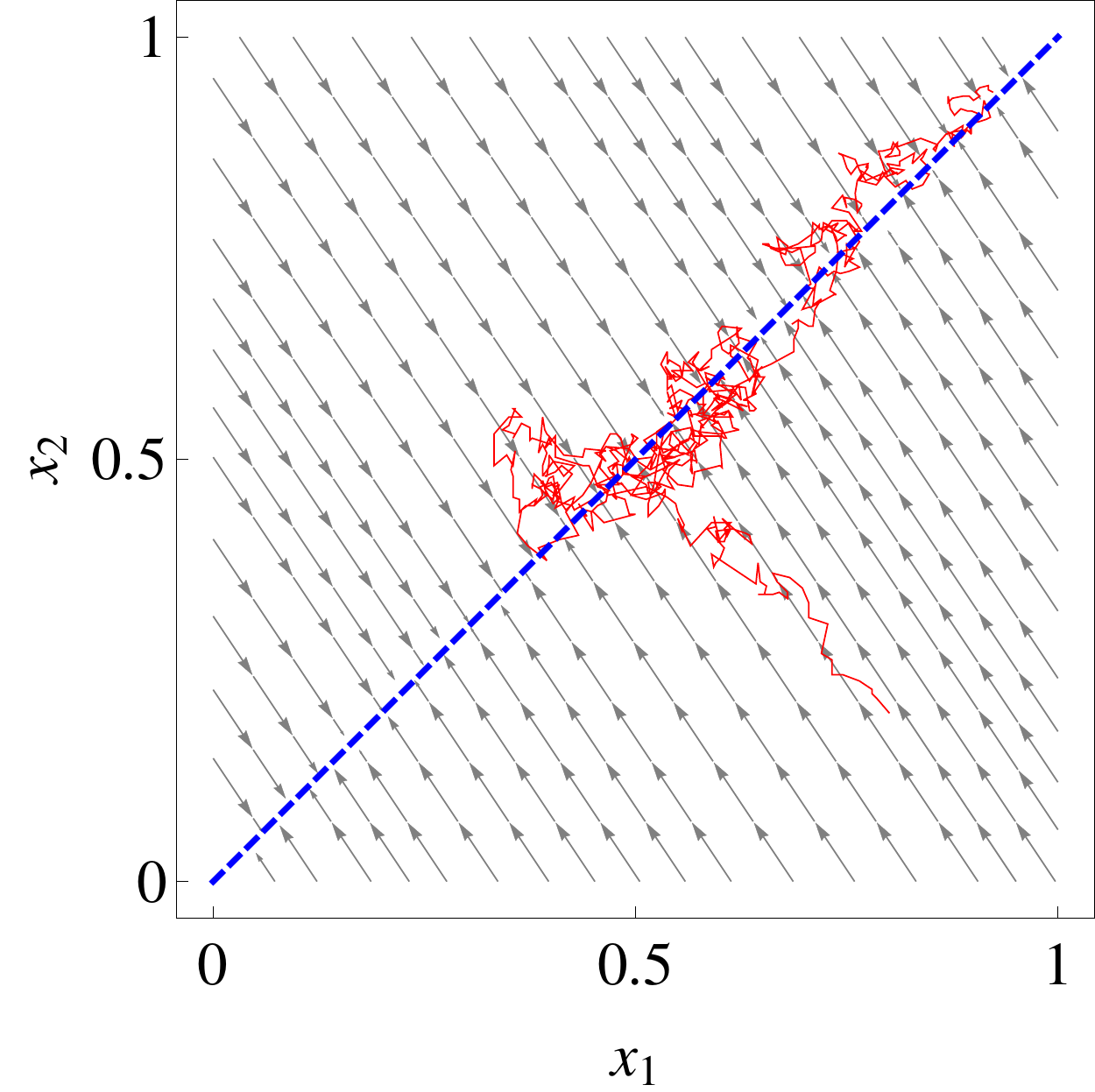}
 \includegraphics[width=0.4\textwidth]{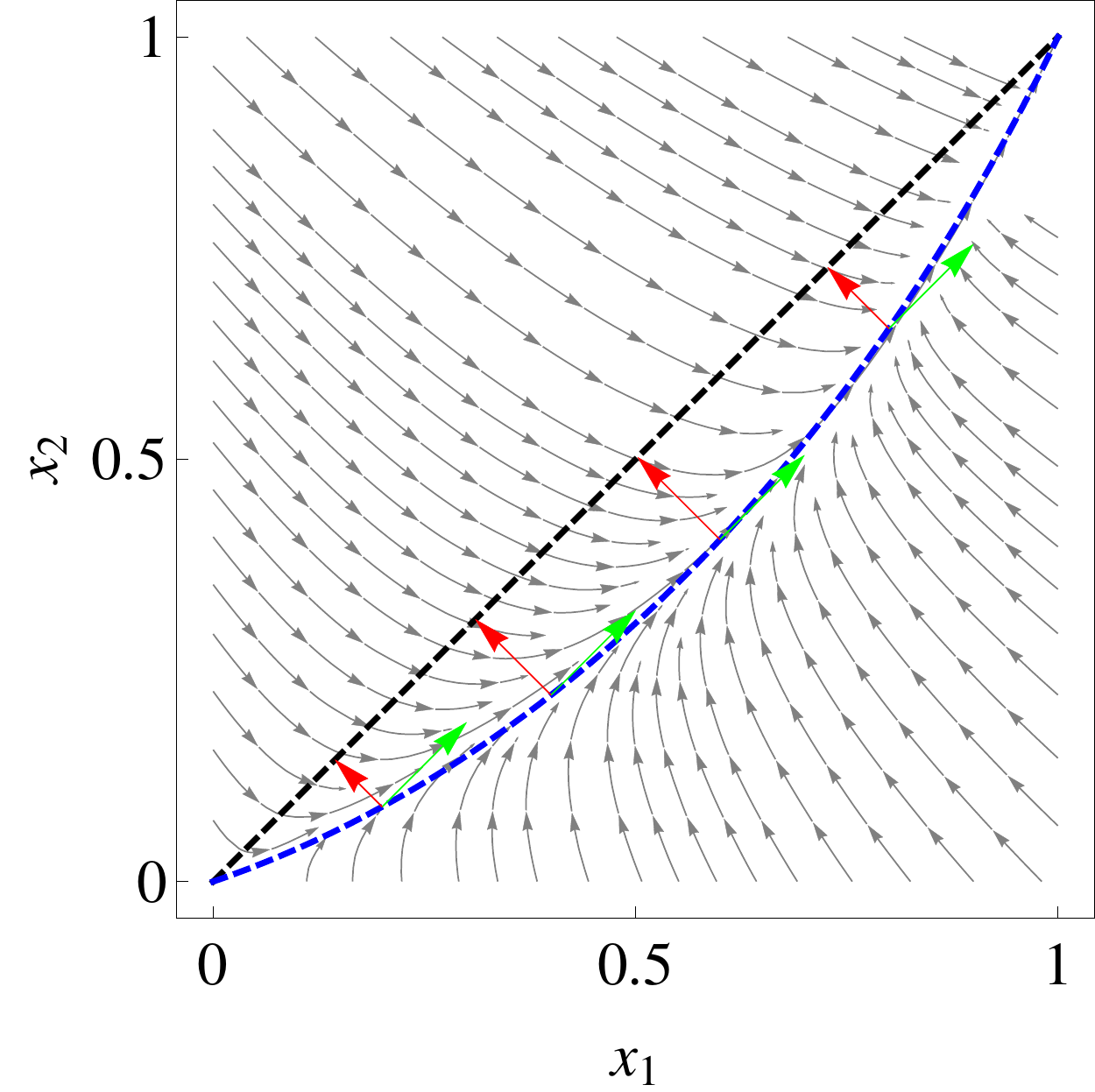}
\caption{Top panel: Time series of an individual stochastic trajectory (red) and deterministic trajectories (grey) for the neutral system, $s=0$. The stochastic trajectory can be seen to quickly collapse onto the deterministic centre manifold $x_{1}=x_{2}$, highlighted in blue, along which stochastic dynamics are observed. Bottom panel: Deterministic trajectories for $s\neq0$ plotted in grey. Here a large value of $s$ has been used ($s=0.1$), in order to emphasise the nature of the dynamics. The order $s$ and $s^{2}$ non-linear terms in \eref{generalSelectionDrift} result in trajectories that are curved relative to the neutral case depicted in the top panel. The variable $z$ is measured along the straight black dashed line, while the approximation of the slow subspace is plotted as a curved blue line. The distance between the slow subspace and the line $\bm{x}=z\bm{v}^{(1)}$ is a function of $s$. The key elements of the approximation are that the system lies on this slow subspace, can move in the direction $\bm{v}^{(1)}$ (green arrows), but cannot move along the directions $\bm{v}^{(2)}$ (red arrows). }
\label{collapse}
\end{figure}

Firstly we consider the neutral deterministic system, that is, we set $s = 0$ and $N \rightarrow \infty$. The system is then linear, entirely governed by the matrix $H$  and hence exactly solvable. The deterministic solution displays a distinct separation of timescales; the system quickly collapses onto a linear subspace, the centre manifold~\cite{wiggins2003}, upon which it remains indefinitely. It is this separation of timescales which we exploit in order to make analytic progress.  If one is only interested in the long-term behaviour of the system, one may effectively ignore this rapid transient and assume that the system reaches the centre manifold very quickly. One can show that this centre manifold is given by $x_{i}=x_{j}$, for all $i,j$ (parallel to the vector $\bm{v}^{(1)}$) independent of the choice of parameters (see \fref{collapse}, upper panel, blue-dashed line). 


Let us continue to consider the neutral system, but now allow the population to be finite so that noise is non-zero. Away from the centre manifold we find that the deterministic dynamics dominate the system's trajectory, dragging it to the vicinity of the centre manifold. Once in this region, the deterministic dynamics cease to dominate and the effect of demographic noise comes into play. This situation is depicted in the upper panel of \fref{collapse}. The deterministic dynamics keep the trajectory of the system confined to a region around the centre manifold, along which it moves stochastically. 

We introduce a variable $z$ which represents the component of the state vector along the centre manifold. The variable $z$ lies on the interval $[0,1]$, so that if $z=0$ then $x_{i}=0$ for all $i$, and likewise if $z=1$, $x_{i}=1$ for all $i$. In order to exploit the separation of timescales in this stochastic setting, we make the following assumptions:
\begin{enumerate}[(i)] 
 \item The system lies on the centre manifold.
 \item The only component of the deterministic dynamics, $\bm{A}(\bm{x})$, is along the centre manifold.
 \item The only component of the noise is in the direction of the centre manifold.
 \item Given a set of initial conditions $\bm{x}_{0}$, the system quickly collapses along a deterministic trajectory to some point, $z_{0}$, on the centre manifold.
\end{enumerate}
This allows us to construct a one-dimensional SDE that describes the dynamics of the system in terms of $z$.

Now what if the system features selection? Firstly the deterministic analogue of the SDE~(\ref{generalSDE}) is now non-linear. However, if $s$ is small (which has already been assumed in the Taylor expansion of \eref{transitionRatesSelection}) one still observes a separation of timescales. The system now quickly collapses to a curved slow subspace rather than a centre manifold. An analytic approximation to the slow subspace is still obtainable, by using a linear approximation for the directions of the transient behaviour. This approximation provides excellent agreement with the slow-subspace observed by numerically simulating the deterministic ODE's, as demonstrated in the bottom panel of \fref{collapse}. As $s \rightarrow 0$ the centre manifold and slow subspace coincide. 

In order to extend the stochastic reduction method to the case with selection, we alter our assumptions slightly, relying in part on the small size of $s$ relative to $H$. Maintaining our definition of $z$ as measuring the distance along $\bm{v}^{(1)}$ (the centre manifold in the neutral case), we make the following approximations:
\begin{enumerate}[(i)] 
 \item The system lies on the slow subspace.
 \item The only component of the deterministic dynamics is in the direction $\bm{v}^{(1)}$.
 \item The only component of the noise is in the direction $\bm{v}^{(1)}$.
 \item Given a set of initial conditions $\bm{x}_{0}$, the system quickly collapses to a point on the slow subspace, whose component in the direction $\bm{v}^{(1)}$ is approximately that of the neutral case, $z_{0}$.
\end{enumerate}
This situation is depicted in the bottom panel of \fref{collapse}; the system moves freely in the direction $\bm{v}^{(1)}$, indicated by green arrows, but does not move in the direction $\bm{v}^{(2)}$, parallel to the red arrows. Again we arrive at a one-dimensional SDE approximating the dynamics along the slow subspace.

The general reduced SDE may be expressed in terms of the variable $z$ as
\begin{equation}
\dot{z} = \bar{A}(z) + \frac{1}{\sqrt{N}}\zeta(\tau) \,, \label{generalSDE1D}
\end{equation}
with 
\begin{equation}
\langle \zeta(\tau)\zeta(\tau') \rangle = \bar{B}(z) \delta(\tau-\tau').
\label{zeta}
\end{equation}
The reduced deterministic term is given by~\cite{projectionPhys}
\begin{eqnarray}
 \bar{A}(z) = s a_{1}  z (1 - z) + s^{2} a_{2} z^{2}(1-z) \nonumber \\ + s^{2} a_{3} z(1-z)(1-2 z) + \mathcal{O}(s^{3})\, , \label{ABar_s_N}
\end{eqnarray}
where parameters $a_{1}$ and $a_{2}$ are defined by
\begin{equation}\label{definea1}
a_{1}  = \sum_{i,j=1}^{\mathcal{D}}u^{(1)}_{i} \frac{G_{ij}}{\beta_{i}}\alpha_{j} \,
\end{equation}
and 
\begin{equation}\label{definea2}
 a_{2} = -  \sum_{i,j=1}^{\mathcal{D}}u^{(1)}_{i} \frac{G_{ij}}{\beta_{i}}\alpha_{j}^{2} ,
\end{equation}
and the parameter $a_{3}$, whose more complicated form arises from contributions on the slow subspace, is
\begin{equation}\label{definea3}
 a_{3} = - \sum_{i,j,k,l=1}^{\mathcal{D}}   u^{(1)}_{i}\frac{G_{ij}}{\beta_{i}}\alpha_{j} \left( \sum_{a=1}^{\mathcal{D}-1} \frac{v^{a+1}_{j}u^{(a+1)}_{k}}{\lambda^{(a+1)}} \right) \frac{G_{kl}}{\beta_{k}}\alpha_{l}    \,. 
\end{equation}
The noise covariance matrix $\bar{B}(z)$, takes the relatively simple form 
\begin{equation}\label{defineBBar}
\bar{B}(z) = 2 b_{1} z (1 - z ) +\mathcal{O}(s)\,,
\end{equation}
where we have introduced the parameter 
\begin{equation}
 b_{1} = \sum_{i,j=1}^{\mathcal{D}}  [u^{(1)}_{i}]^{2}G_{ij}\beta_{i}^{-2}   \,.
\label{defineb1}
\end{equation}

Finally, the initial condition for the one-dimensional system must be defined mathematically. Isolating the component of the initial state vector, $\bm{x}_{0}$, along the centre manifold~\cite{projectionPhys}, one finds
\begin{eqnarray}
 z_{0}=\sum_{i=1}^{\mathcal{D}}u^{(1)}_{i}x_{0i} \, .
\end{eqnarray}
These expressions completely define the reduced system.

Our approximate system effectively ignores initial transient dynamics under the assumption that these are negligible when asking questions about the long-time properties of the system. The reduced system is therefore ideally suited to answering questions about fixation, which will occur on a much slower timescale than it takes for the system to reach the centre manifold or slow subspace. We expect that the reduced system will provide a good approximation to the full system so long as the linear deterministic dynamics are dominant. This has been found to be the case so long as the magnitudes of the real parts of the non-zero eigenvalues of $H$ are greater than $N^{-1/2}$ and $s$, as discussed in \cite{projectionPhys}.

\section{Analysing the reduced model}\label{secApplications}

It has been stated in the previous section that, given the nature of the fast-timescale approximation, the reduced system is suited to answering questions about fixation. In order to test the predictions of the reduced model against stochastic simulations of the full model defined by \eref{transitionRatesSelection}, we choose to use the probability of fixation and time to fixation as metrics.

We begin by recalling that the It\={o} SDE \eref{generalSDE1D} is entirely equivalent to the FPE
\begin{eqnarray}\label{generalFPE1D}
 \frac{\partial p(z,t)}{\partial t} = &-& \frac{1}{N}\frac{\partial}{\partial z} \left[\bar{A}(z)p(z,t)\right] \nonumber \\ &+& \frac{1}{2N^{2}}\frac{\partial^{2}}{\partial z^{2}} \left[\bar{B}(z)p(z,t)\right] \,.
\end{eqnarray}
We can therefore calculate, just as in the well-mixed case discussed in Section
\ref{secNeutralReview}, the probability of fixation, $Q(z_{0})$ and the time to fixation $T(z_{0})$ by solving \eref{BFPQ} and \eref{BFPT} with $x$ replaced by $z$ and using the reduced terms $\bar{A}(z)$ and $\bar{B}(z)$ given by \eref{ABar_s_N} and \eref{defineBBar}.

\subsection{Fixation probability and mean fixation time in the neutral case}\label{secProbabilitiesNeutral}

To obtain the neutral reduced model we simply set $s=0$ in \eref{generalSDE1D}, which yields $\bar{A}(z)=0$ by \eref{ABar_s_N}, with $\bar{B}(z)$ given by \eref{defineBBar}. The reduced equation has the same functional form as that of the well-mixed case, \eref{1Ddiffusion}, but with $N^{2}$ scaled by $b_1$. Thus the calculations of the fixation probability, $Q(z_{0})$, and fixation time, $T(z_{0})$, follow in a straightforward manner:
\begin{eqnarray}
  Q(z_0) &=& z_0, \label{Q_s_0} \\
  T(z_{0}) &=& - (N_{\rm Tot}r_{N})^2\left[ (1-z_{0})\ln{(1 - z_{0})} \right. \nonumber \\ & & \hspace{3.5cm} \left. +  z_0 \ln{(z_0)} \right] \, ,   \label{T_s_0}
\end{eqnarray}
where
\begin{eqnarray}\label{defineRN}
N_{\rm Tot}=N\sum_{k=1}^{\mathcal{D}}\beta_{k} \quad \mathrm{and} \quad r_{N} = \left( \sqrt{b_1}\sum_{k=1}^{\mathcal{D}}\beta_{k} \right)^{-1} \, . 
\end{eqnarray}
These are identical to the results in the well-mixed case if in Eqs.~(\ref{1D_Q_s_0}) and (\ref{1D_T_s_0}) we replace $x_0$ by $z_0$ and replace $N$ by $N_{\rm Tot}r_{N}$.

The parameter $N_{\rm Tot}$ is the total size of the unstructured population. A natural interpretation is that the population behaves as a well-mixed population with a new `effective population size' $r_{N}N_{\rm Tot}$. We will however avoid this terminology, since the variable $z$ is not directly equivalent to the allele frequency within the global population and, in addition, since it is frequently used in other situations in which its meaning differs from that which we would ascribe to it~\cite{whitlock1997,ewens2004}. 

In agreement with the results of Nagylaki \cite{nagylaki1980SM} (whom we recall considered a Wright-Fisher migration model with non-overlapping generations), the value of $r_{N}$ takes rather a simple form in the case that migration rate matrix, $G$, is symmetric~\cite{projectionPhys}. Direct substitution using Eq.~({\ref{defineH}) and the symmetry of $G$ shows that $\sum^{\mathcal{D}}_{i=1} \beta_i H_{ij} = 0$ for all $j$, and so $\beta_i$ is the left-eigenvector of $H$ with eigenvalue $\lambda_{1}=0$. So $\beta_i$ must be proportional to $u_i$, and since the normalisation of $u_i$ has already been fixed through Eq.~(\ref{orthonormal}) and the choice of $\bm{v}^{(1)}$, we have that $u^{(1)}_{i} = \beta_{i}(\sum_{k=1}^{\mathcal{D}}\beta_{k})^{-1}$. Substituting this expression into Eq.~(\ref{defineb1}) and using $\sum_{i,j}G_{ij}=1$, gives $r_{N}=1$. When the migration is symmetric therefore, the population behaves on the same timescale as a well-mixed population of equal size, albeit with a weighted initial allele frequency, $z_{0}$.

Our results diverge from those of Nagylaki outside this limit however. In his model and analysis, it was found that $r_{N} \leq 1$, whereas we find no strict upper bound on the value of $r_{N}$. Indeed, in \Sref{secHub} we will find that $r_{N}$ may be significantly higher than one in some particular situations. This serves to emphasise that $r_{N}N_{Tot}$ does not provide an effective population size, but rather describes a typical timescale for fixation.

To demonstrate the range of values $r_{N}$ can take, a numerical study can be conducted. An ensemble of random migration matrices, $m$, are first generated. We have to be careful to pick the elements of $m$ such that the normalisation condition $\sum_{i=1}^{\mathcal{D}}m_{ij} =1$ holds. Additionally, from a modelling perspective, we would like to see the diagonal elements of $m$ larger than $1/2$ at least, $m_{ii}>1/2$, so that the probability of an offspring not migrating is greater than the probability it migrates. For each random migration matrix generated, an $r_{N}$ may then be calculated to give an indication of a potential distribution of $r_{N}$ values. Since we expect the reduction technique to become unreliable if any of the real parts of the non-zero eigenvalues of $H$ are smaller in magnitude than $N^{-1/2}$ (see \Sref{secReduction} and~\cite{projectionPhys}), we discard any $m$ matrices that yield such values.

Initially we consider systems with $\mathcal{D}=4$ and $\beta_{i}=1$ for all $i$. The values for $r_{N}$ are plotted in a histogram in \fref{rHistograms}; while no strict upper value for $r_{N}$ exists, the distribution in this parameter regime does not allow for $r_{N}>1$. We note however, that this is a feature of our modelling choice; if we remove the restriction $m_{ii}>1/2$, $r_{N}$ can take a range of values around one. Further, if we allow the island sizes to vary (as in \fref{rHistograms}, inset) the distribution of $r_{N}$ values is altered to allow $r_{N}>1$. 

We can test these predictions against a Gillespie simulation of the full system, defined by the transition rates \eref{transitionRatesNeutral1} and \eref{transitionRatesNeutral2}. The time to extinction of one or other of the alleles, $T(z_{0})$, calculated from the effective theory, \eref{T_s_0}, is used as a measure of the timescale of the effective system. We find excellent agreement between simulation and theory across a range of parameters, as shown in the bottom panel of \fref{rHistograms}. 


\begin{figure}[h!]
\centering
\includegraphics[width=0.45\textwidth]{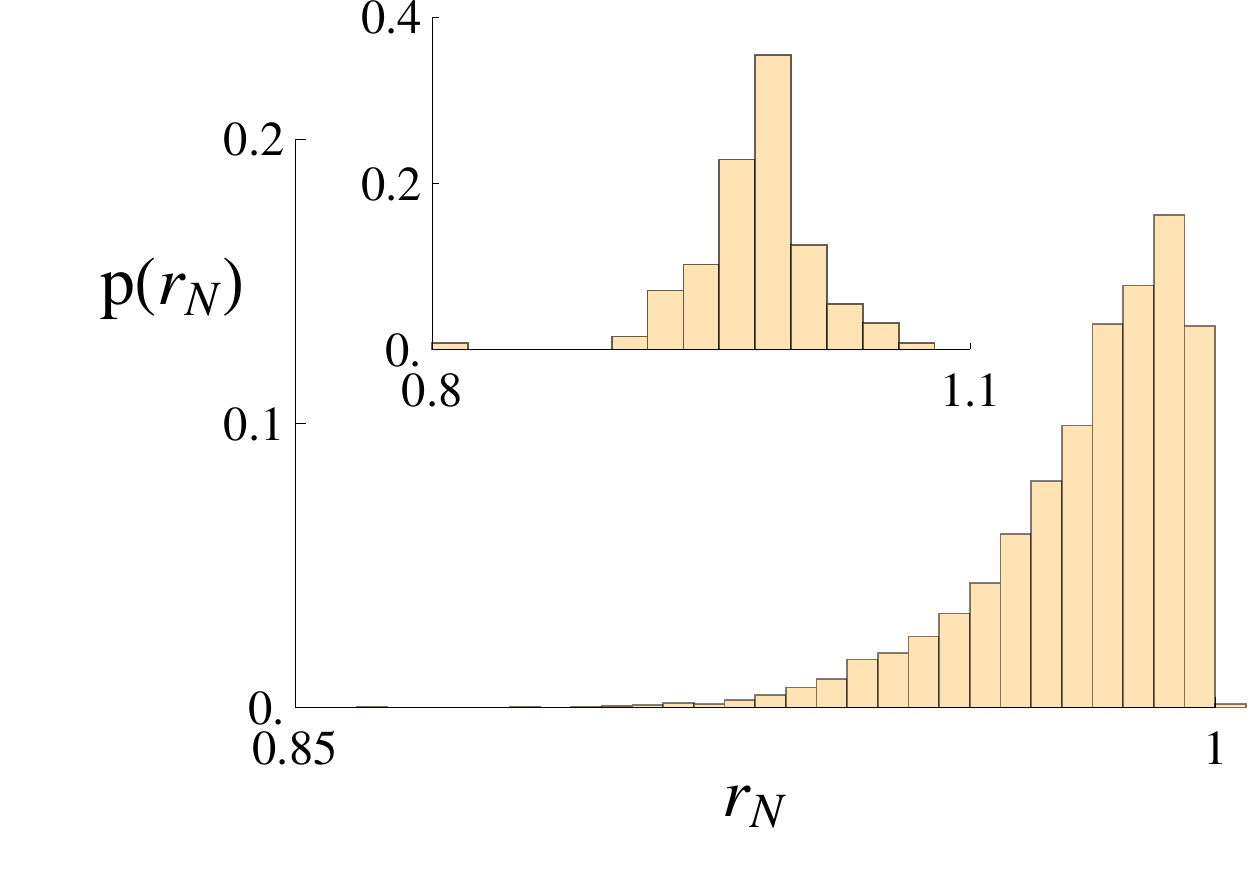}
\includegraphics[width=0.45\textwidth]{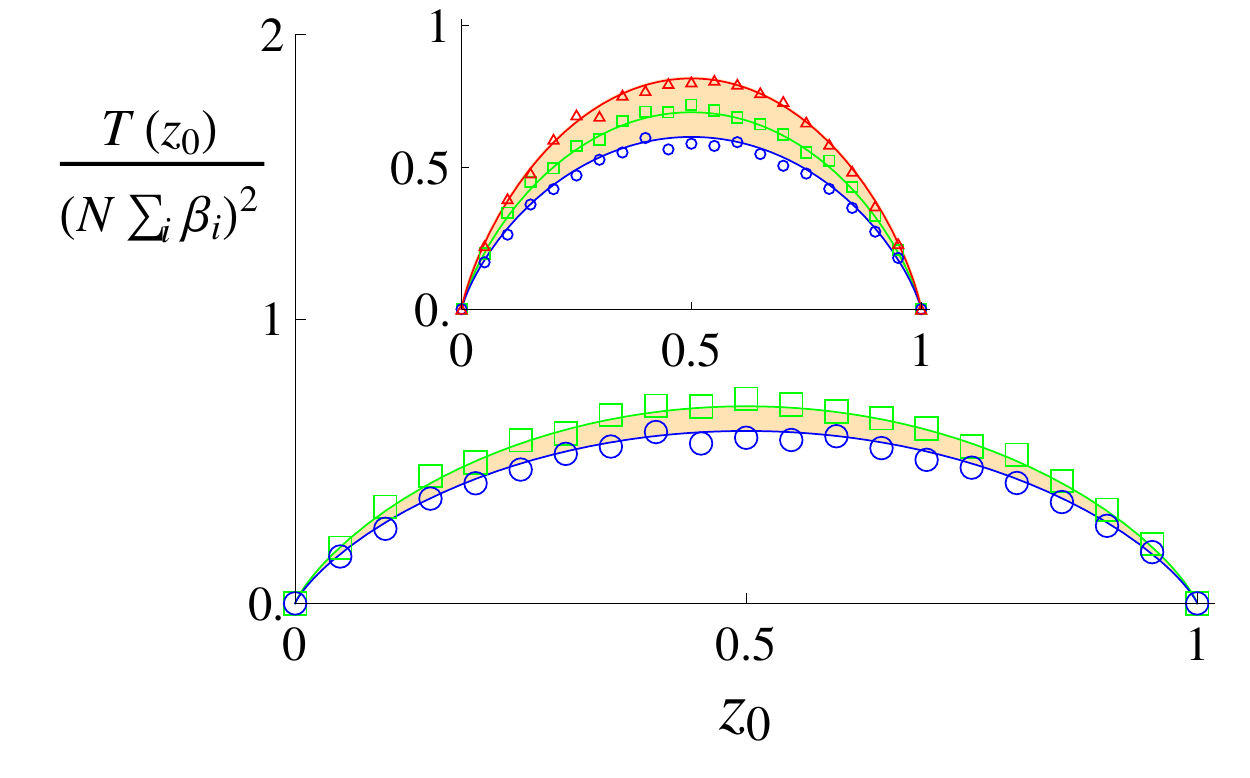}
\caption{Top panel: Histograms of values of $r_{N}$ obtained for a $\mathcal{D}=4$ system with randomly generated and appropriately normalised migration matrices, $m$, and $s=0$. Main histogram shows results obtained when all island sizes are the same, $\beta_{i}=1$ for all $i$. The inset histogram is obtained when the elements $\beta_{i}$ are themselves random integers. We have taken $N=250$ in both cases, and hence have discounted any of the random systems that yield a non-zero eigenvalues with a real part greater than $-N^{-1/2}$. Bottom panel: Plots of the mean time to fixation as a function of the initial condition $z_{0}$. Analytic predictions from the reduced model, \eref{T_s_0}, are plotted as continuous lines, while the results from simulation are plotted as symbols. Plots in blue/circles correspond to the smallest $r_{N}$ values obtained in the histograms (top panel), those in green/squares are obtained from  systems with symmetric $G$ matrices ($r_{N}=1$), and those in  red/triangles correspond to the largest $r_{N}$ values obtained in the histograms. Once again in the main graph all islands are of the same size, while in the inset plot $\beta_{i}$ is allowed to vary.}
\label{rHistograms}
\end{figure}


\subsection{Fixation probability and mean fixation time in the case with selection}\label{secProbabilitiesSelection}

Once again we seek to solve \eref{BFPQ} and \eref{BFPT} with $x$ replaced by $z$ and $\bar{A}(z)$ and $\bar{B}(z)$ given by \eref{ABar_s_N} and \eref{defineBBar}, but now with some selective bias for one or other of the alleles, such that $s\neq0$. We begin by noting that \eref{ABar_s_N} can be written more compactly as 
\begin{equation}\label{ABar_s_N_compact}
 \bar{A}(z_{0}) = s z_{0}(1 - z_{0})(k_{1} - s k_{2} z_{0}) \, ,
\end{equation}
with 
\begin{equation}
 k_{1} = a_{1} + s a_{3} \, \quad \mathrm{and} \quad k_{2} = 2 a_{3} - a_{2} \,.
\end{equation}

We can now solve \eref{BFPQ} to obtain an expression for the probability of fixation. We shall merely state the result here; full details of the calculation are given in \ref{appSol}. Defining the function
\begin{equation}
 l(z_{0})= \sqrt{\frac{N}{(2b_{1} |k_{2}|)}}(s k_{2} z_{0} - k_{1}),
\end{equation}
the probability of fixation, given initial weighted frequency of $A$ allele $z_{0}$, is given by
\begin{equation}
Q(z_0) = \frac{ 1 - \chi(z_0)}{1 - \chi(1)}; \ \ \ \ 
\chi(z_0) = \frac{f(l(z_0))}{f(l(0))}, 
\label{Q_s_N}
\end{equation}
where the form of the function $f$ depends on the sign of $k_2$. If $k_{2}<0$ the function $f$ is the complementary error function~\cite{handbook1965}, if $k_{2}<0$, it is the imaginary error function ~\cite{HTFS}.

The form of $Q(z_{0})$ is more complex as compared to the neutral case, for which we found that the metapopulation model behaved analogously to the well-mixed model (see Eq.~(\ref{Q_s_0}) and Eq.~(\ref{1D_Q_s_0})). However, we can gain further insight into the model by considering broadly two different parameter regimes. First, we examine the situation in which the advantageous allele is the same on each of the demes. In this case we find that the parameters $a_{1}$, $a_{2}$ and $b_{1}$ are all of order $1$, while we have taken $N$ large. Given that the function $l(z)$ is then relatively large, we can perform an asymptotic expansion of the error function (see \ref{appSol}, \eref{asymp_erfc} and \eref{asymp_erfi}) to find that for both $k_2 < 0$ and $k_2 > 0$,
\begin{eqnarray}
\hspace{-0.6cm} \chi(z_0) \approx  \left( 1 - \frac{k_2}{k_1} s z_0 \right)^{-1}  \exp\left( - \frac{k_1}{b_1} sNz_0 + \frac{k_2}{2b_1} s^2 N z_0^2 \right) \nonumber . 
\end{eqnarray}
Having obtained this expression, valid for large $l(z_{0})$, we can make a further approximation for small $s$. Taking only linear $s$ terms from the above equation, we obtain
\begin{eqnarray}
 \chi(z_{0}) &\approx& \exp \left( - \frac{a_{1}}{b_{1}}  s N z_{0} \right) , \label{Q_s_N_Asymp}
\end{eqnarray}
which is the form given in \eref{1D_Q_s_N}, but with a selection strength (or system size) weighted by the ratio $a_{1}/b_{1}$. One may also obtain this solution by simply truncating the reduced term \eref{ABar_s_N} at order $s$ rather than $s^{2}$, and solving the relevant ODE, \eref{BFPQ} in Section \ref{secNeutralReview}. We find that this provides a very good approximation in this regime, as demonstrated in \fref{QTAssymp}.


\begin{figure}[h!]
\includegraphics[width=0.45\textwidth]{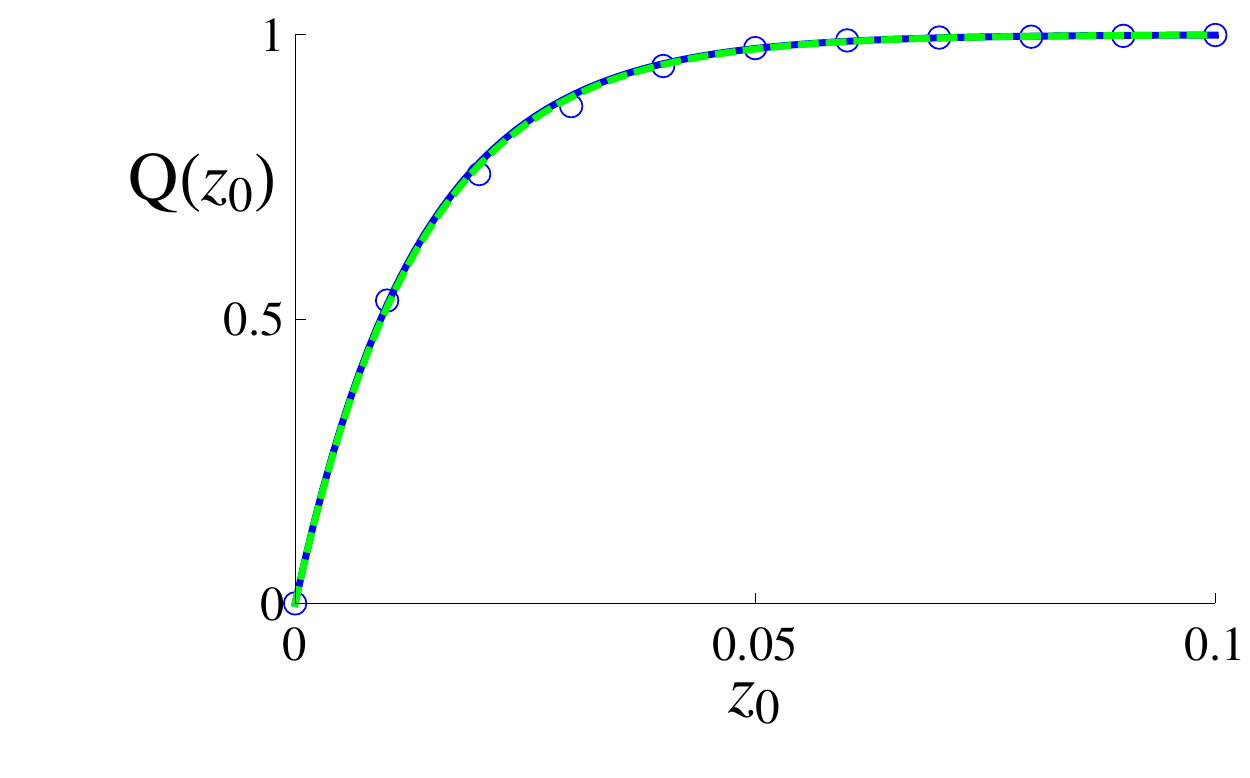}
\includegraphics[width=0.45\textwidth]{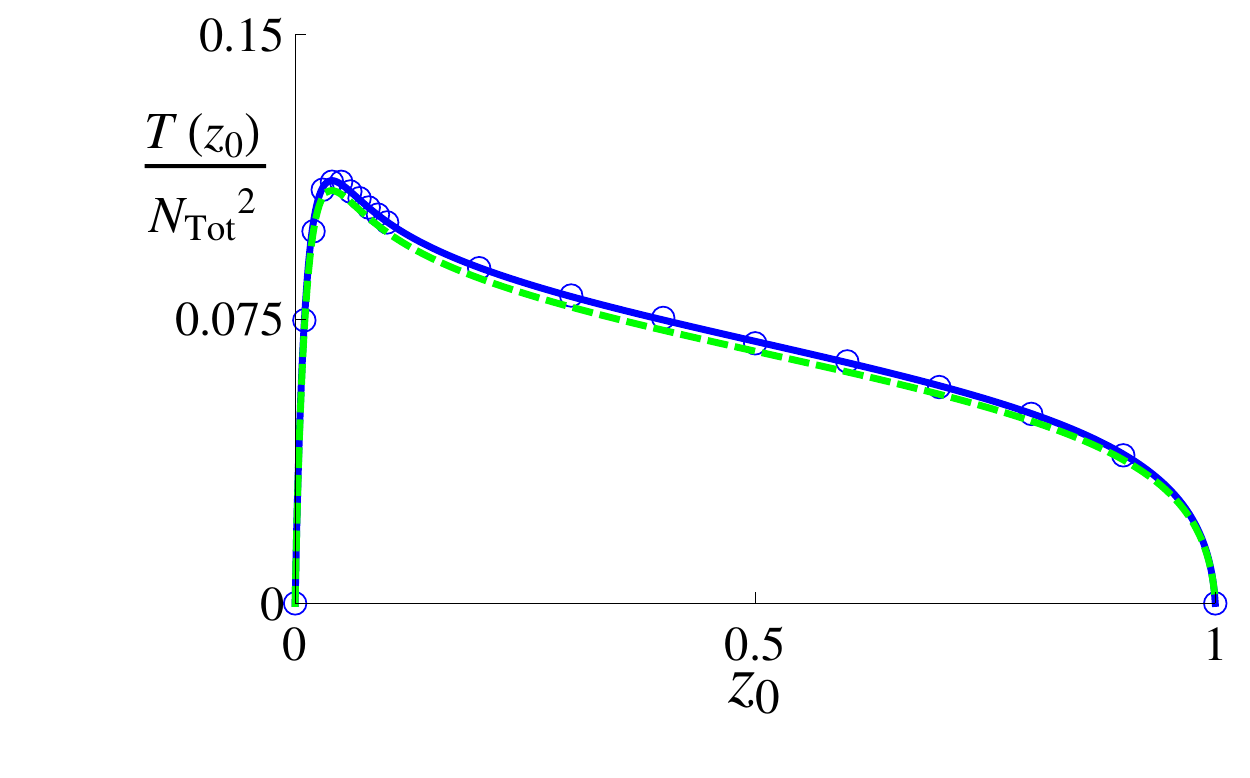}
\caption{Plots for the probability of fixation, $Q(z_{0})$, at low values of $z_{0}$, and the mean time to fixation, $T(z_{0})$, in a system where $s$ is of order $N^{-1/2}$. Continuous blue lines are obtained from the reduced model (using \eref{Q_s_N} with $k_{2}>0$ for $Q(z_{0})$ and solving \eref{BFPT} with $\bar{A}(z)$ given by \eref{ABar_s_N} and $\bar{B}(z)$ by \eref{defineBBar}, numerically for $T(z_{0})$). Parameters used here are $\mathcal{D}=4$, $s=0.05$, $N=400$, $\bm{\beta}=(1,1,1,2)$, $\bm{\alpha}=(1,0.1,0.5,1)$. We omit the explicit form of the migration matrix here for brevity. Since there are no demes in which selection acts in a contrary direction to any of the others, we can use the asymptotic expansion for $Q(z)$, \eref{Q_s_N} with \eref{Q_s_N_Asymp}. The asymptotic expression is plotted by a green dashed line;  it is indistinguishable from the full order $s^{2}$ solution in this regime. For $T(z_{0})$ we also plot the  first order in $s$ solution as a green dashed line; while qualitatively similar to the full solution, there is some numerical discrepancy.}
\label{QTAssymp}
\end{figure}


If the direction of selection varies from deme to deme however, then from a consideration of the forms of $a_{1}$ and $a_{2}$ one can see that there may be some cancellations. This reduces the size of these parameters and invalidates the use of the asymptotic expansion; one must therefore resort to evaluating the expressions given in Eq.~(\ref{Q_s_N}) numerically. We find excellent agreement across a wide range of parameters, as shown in \cite{projectionPhys}.

One may also calculate the mean time to fixation, $T(z_{0}$) from \eref{BFPT} with $x_{0}$ replaced by $z_{0}$. There are singular points of the differential equation at the boundaries, and care is required when imposing boundary conditions. These aspects are discussed in \ref{appMTF}, where we find expressions for $T(z_{0})$ in terms of well-defined integrals at various order of $s$. For instance, to first order in $s$ we find that
\begin{eqnarray}
 \hspace{-0.5cm} T(z_0) =&& \hspace{-0.5cm} c_{2}\,\left[ 1 - e^{-M\sigma z_{0}} \right] \nonumber \\
         &&\hspace{-0.5cm} - M^2 e^{-M\sigma z_0} \int^{z_0}_{0} dx\,e^{M\sigma x}\,\left[ \ln x - \ln (1-x) \right]\, , \nonumber \\
\label{T_linear_in_s}
\end{eqnarray}
where
\begin{equation}
c_{2} = \frac{M^2 e^{-M\sigma}}{1 - e^{-M\sigma}}\,\int^{1}_{0} dx\,e^{M\sigma x}\,\left[ \ln x - \ln (1-x) \right].
\label{c_2}
\end{equation}
Here $M = N/\sqrt{b_1}$ and $\sigma = a_{1}s/\sqrt{b_1}$. The integrals in Eqs.~(\ref{T_linear_in_s}) and (\ref{c_2}) may be expressed as combinations of the exponential integral function~\cite{handbook1965} and logarithms, but they may also be easily evaluated numerically. At second order in $s$ the results are more complex, but can again be straightforwardly evaluated. Once again we find very good agreement between the reduced model and simulation (see \fref{QTAssymp} and \cite{projectionPhys}).

We now proceed to discuss how some of the predictions of the reduced system relate to the behaviour of the full system. In \Sref{secMigrationSeletionBalance} we will begin by discussing a case in which the system admits a polymorphic equilibrium in the deterministic limit. We will then consider the case of a stylised network topology, the hub, in order to both illustrate the details of the method and to reveal how the partitioning of the population into demes can significantly alter the behaviour of the system.

\subsection{Migration-selection balance}\label{secMigrationSeletionBalance}

So far we have introduced our migration model and applied the reduction technique to arrive at a one-dimensional SDE or FPE which captures the dynamics of the metapopulation. Having obtained these general results, it is now both interesting and instructive to consider a specific system. Of particular note, is the case in which the deterministic system, $N\rightarrow \infty$, predicts a migration-selection balance of the two alleles; both alleles $A$ and $B$ can coexist in a stable polymorphic equilibrium.

Let us begin be considering the deterministic equations, \eref{generalSDE} with $\bm{A}(\bm{x})$ given by \eref{A_second_order} and $N\rightarrow \infty$. For clarity we restrict our attention to a two island system with equal island sizes, $\mathcal{D}=2$, $\bm{\beta}=(1,1)$, and a symmetric migration matrix
\begin{eqnarray}
 m = \left( \begin{array}{cc} \theta       & (1-\theta)  \\
                        (1-\theta) &    \theta    \\ \end{array} \right) \, , \label{migBalanceM}
\end{eqnarray}
parametrised by the appropriately normalised probability of offspring \emph{not} migrating, $0<\theta<1$. While the behaviour of the linear neutral system was straightforward, the introduction of the non-linear $s$ terms in \eref{ABar_s_N} allows for more complicated behaviour. One finds that for $s\neq0$, nine fixed points emerge. Two of these are at the points of fixation $\bm{x}^{*}_{A}=(1,1)$ and $\bm{x}^{*}_{B}=(0,0)$. While a numerical analysis finds that six of the remaining seven fixed points have values outside the physical range, a final fixed point, $\bm{x}^{*}_{PE}$, may arise in between $\bm{x}=(0,0)$ and $\bm{x}=(1,1)$, under the condition that the selective pressure works in opposite directions on each of the demes. Further, one can observe that only one of the fixed points $\bm{x}^{*}_{A}$, $\bm{x}^{*}_{B}$, or $\bm{x}^{*}_{PE}$ is stable for a given set of parameters. An overview of the situation is given in \fref{figMigrationSelectionBalance1} (top panel); while the region of stable polymorphic equilibrium may appear large in this highly symmetric parameter regime, we note that in general it only occurs for a very restricted range of parameters.


\begin{figure}[h!]
\includegraphics[width=0.45\textwidth]{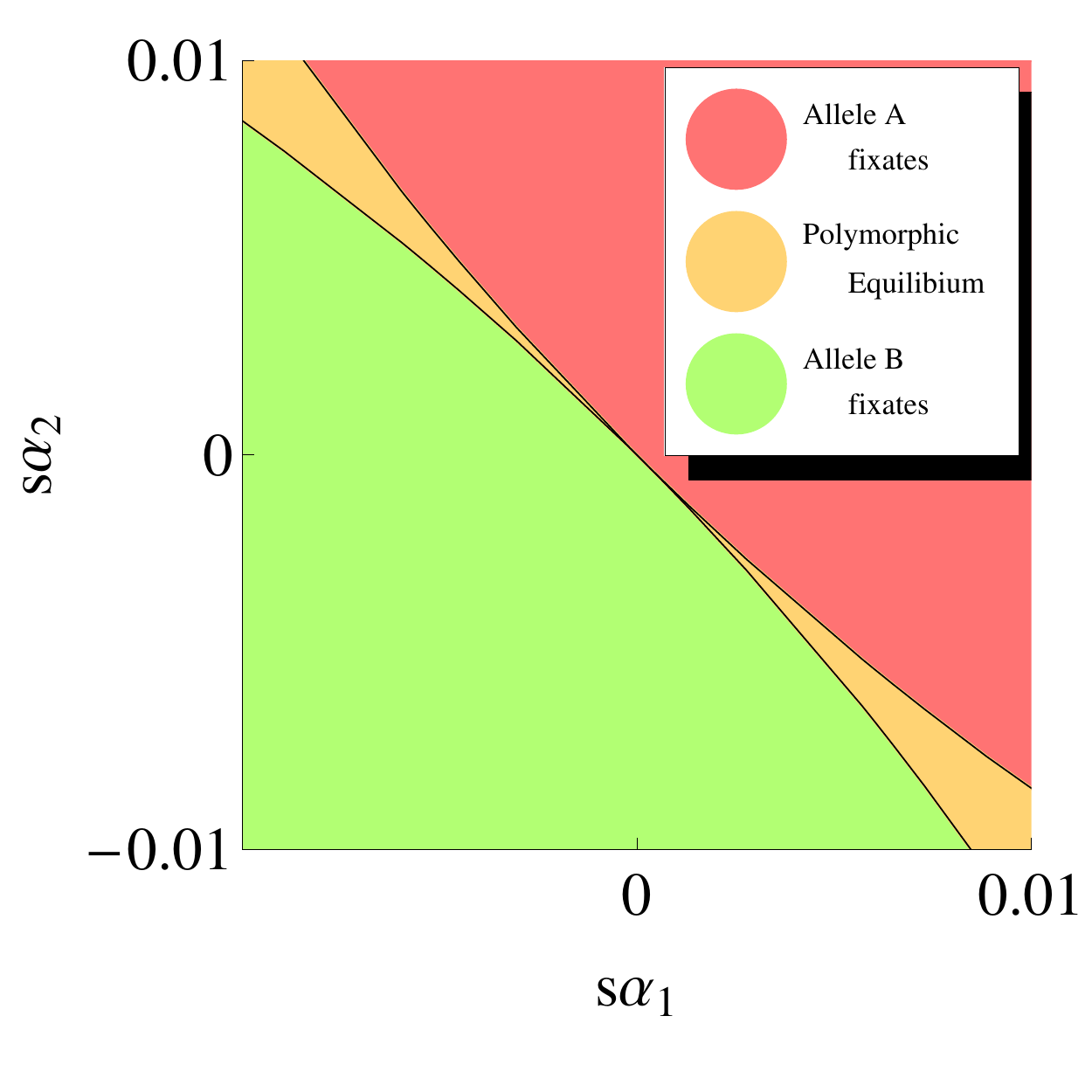}
\includegraphics[width=0.45\textwidth]{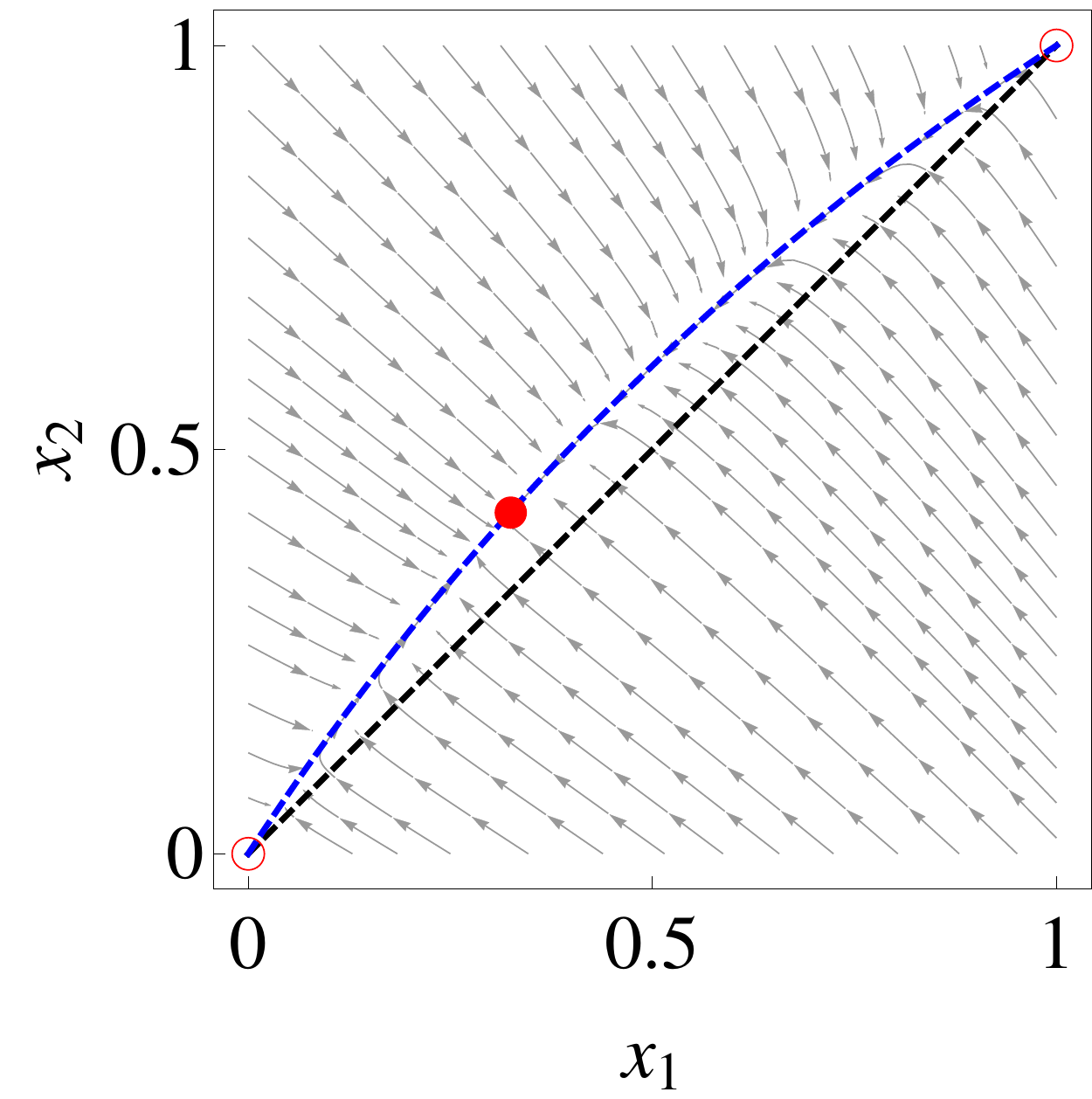}
\caption{Top panel: Plot of regions of stability for the fixed points $\bm{x}^{*}_{A}$ (red, upper-right region), $\bm{x}^{*}_{B}$ (green, lower-left region) and $\bm{x}^{*}_{PE}$  (orange, central region), for a system $\mathcal{D}=2$, $\bm{\beta}=(1,1)$ and $m$ given by \eref{migBalanceM} with $\theta=0.95$. Bottom panel: Deterministic trajectories (grey) for the same system with $\theta=0.8$, $\bm{\alpha}=(1,-1)$ and $s=0.14$. The stable fixed point $\bm{x}^{*}_{PE}$ is indicated by a red disc, while unstable fixed points, $\bm{x}^{*}_{A}$ and $\bm{x}^{*}_{B}$, are red circles. The straight line $\bm{x}=z\bm{v}^{(1)}$ is plotted as a black dashed line, while the analytic approximation of the curved slow subspace is plotted as a blue dashed line. The location of $\bm{x}^{*}_{PE}$ directly on the approximate slow subspace serves to further emphasise the quality of the approximation.}
\label{figMigrationSelectionBalance1}
\end{figure}
 

Let us now restrict our attention to a perfectly symmetric set of parameters by setting $\bm{\alpha}=(1,-1)$. A phase diagram for this system is shown in \fref{figMigrationSelectionBalance1} (bottom panel). It is interesting to note the position at which the fixed point $\bm{x}^{*}_{PE}$ is found. One might expect, given the highly symmetric nature of the system, that it would be found equidistant between the points of fixation of allele $A$ and allele $B$, $\bm{x}^{*}_{A}$ and $\bm{x}^{*}_{B}$. While this is true at first order in $s$, at second order $\bm{x}^{*}_{PE}$ is shifted closer to $\bm{x}^{*}_{B}$. Further, the stability of this fixed point increases with increasing $s$. We may now ask, how does this deterministic behaviour in such a regime impact the predictions of the reduced stochastic system, \eref{generalSDE1D}?

Firstly we note that to first order in $s$, our deterministic term $\bar{A}(z)$ in \eref{ABar_s_N}, admits no fixed point other than $z=0$ and $z=1$. We would expect, however, that the first order in $s$ description would work well for particularly small values of $s$, say $s\approx 1/N$. Indeed, this is what we find for small $s$; the deterministic drive towards the polymorphic fixed point is sufficiently weak that its existence has little effect on the probability of fixation or mean time to fixation. The probability of fixation is then well approximated by \eref{Q_s_N_Asymp} and the time to fixation by \eref{BFPT} with \eref{ABar_s_N} to first order in $s$, as seen in \fref{figMigrationSelectionBalance2} (green/square plot).


\begin{figure}[h!]
\includegraphics[width=0.45\textwidth]{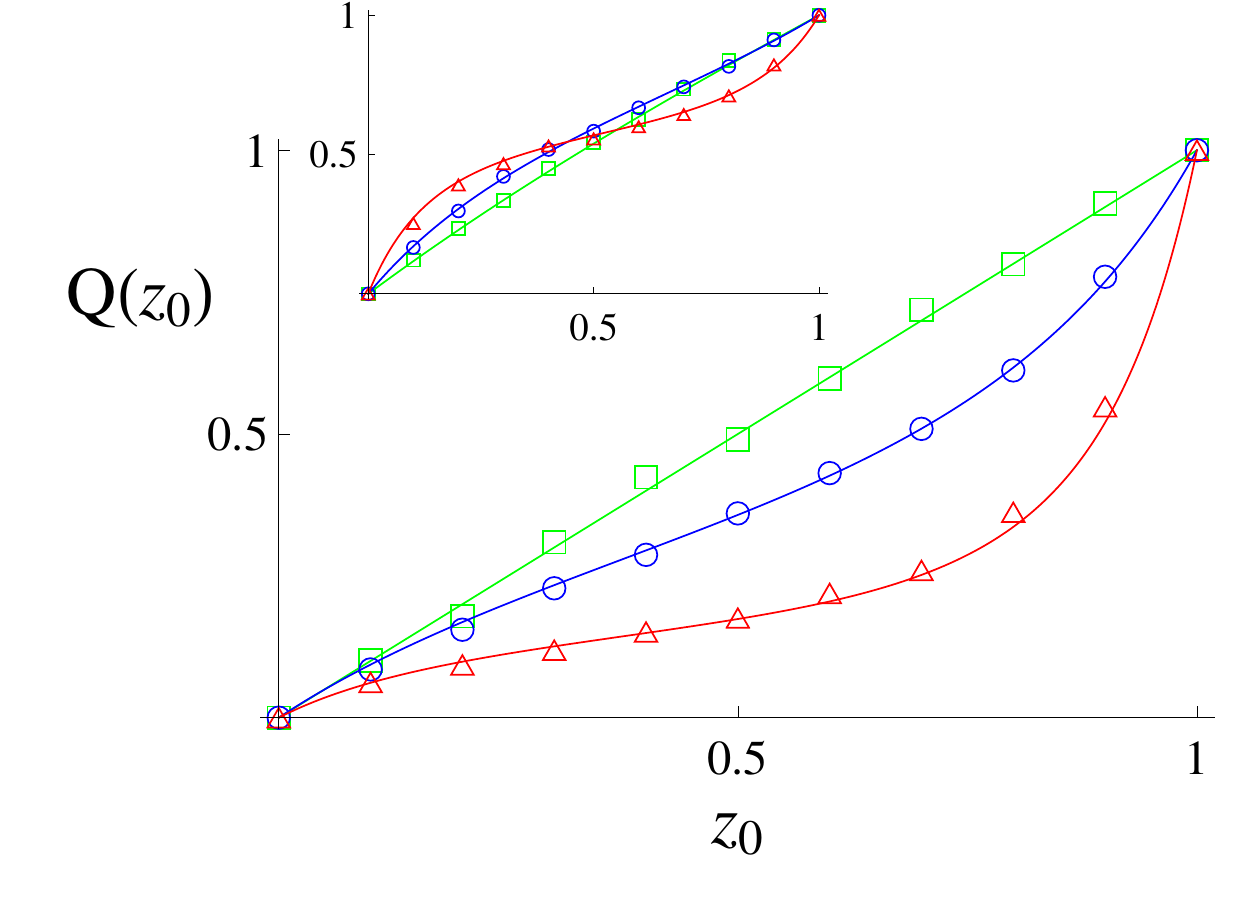}
\includegraphics[width=0.45\textwidth]{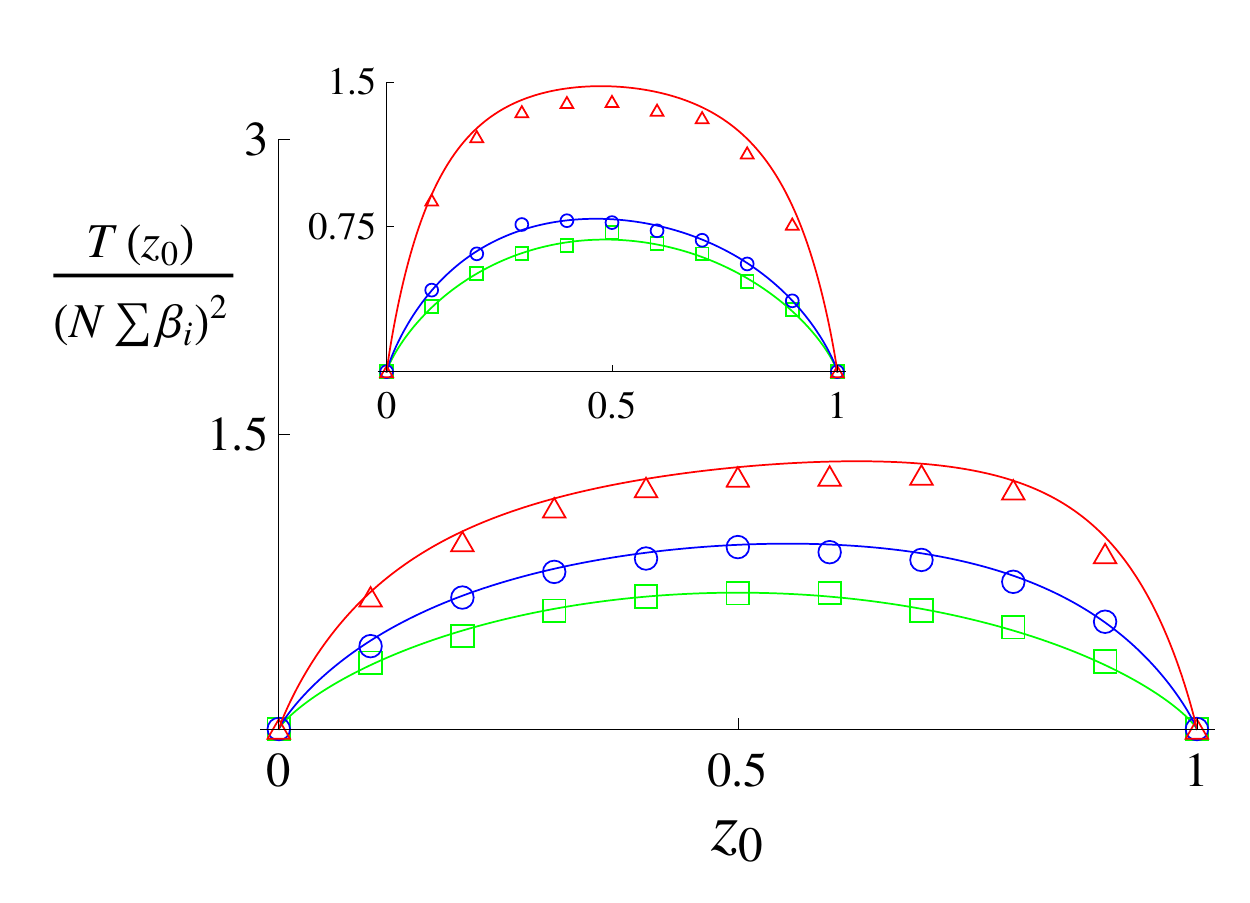}
\caption{Plots of the probability of fixation, $Q(z_{0})$, and mean time to fixation, $T(z_{0})$, as a function of the projected initial conditions, for systems featuring symmetric migration balance. Main plots feature $\mathcal{D}=2$, $\bm{\beta}=(1,1)$, $\bm{\alpha}=(1,-1)$ and $m$ given by \eref{migBalanceM} with $\theta=0.8$ for increasing values of $s$. Continuous lines are obtained from the reduced one-dimensional model, while symbols are obtained from stochastic simulation with $N=300$. Plots in green/squares correspond to $s=1.66\times10^{-3}$ with $Q(z_{0})$ and $T(z_{0})$ obtained from a first order solution to \eref{BFPQ} and \eref{BFPT} (see \eref{Q_s_N_Asymp}). The remaining plots are calculated from second order solutions to \eref{BFPQ} and \eref{BFPT} (see \eref{Q_s_N}) with $s=5.7\times10^{-2}$ for the blue/circle plots and $s=0.11$ for the red/triangle plots. Inset plots meanwhile, are  obtained from parameters $\mathcal{D}=5$, $\bm{\beta}=(1,1,1,1,2)$, $\bm{\alpha}=(1,1,2.2,-1,-1)$ and an $m$ matrix with an unspecified structure. }
\label{figMigrationSelectionBalance2}
\end{figure}


For larger values of $s$, the stability of the polymorphic fixed point increases in the deterministic limit. To capture the effect on $Q(z_{0})$ and $T(z_{0})$ one must solve \eref{BFPQ} and \eref{BFPT} to second order in $s$ (using \eref{ABar_s_N} in full). One finds the probability of fixation begins to `plateau' across a range of initial conditions as $s$ increases, with the fixation of allele $B$ becoming increasingly likely. This counter-intuitive break in symmetry can be viewed as a consequence of the skewed fixed point, which biases the system towards fixation at $\bm{x}=(0,0)$. The reduced model captures the behaviour extremely well, as observed in \fref{figMigrationSelectionBalance2}, top panel. The mean time to fixation meanwhile begins to increase, diverging as the deterministic fixed point holds the system in its vicinity for longer and longer. For these very large, arguably unphysical values of $s$, the reduced system begins to over-predict the rapidly increasing time to fixation, as seen in \fref{figMigrationSelectionBalance2}. This can also be seen as a consequence of $s$ becoming larger than $|\Re{(\lambda^{(2)})}|$.

In this section we have focused on a very restricted set of parameters to illustrate the effect of migration-selection balance. While such stable polymorphic equilibria clearly exist for a host of other parameters, including multiple islands of differing sizes and various selection pressures, the parameter range in which they exist becomes increasingly small relative to the full parameter space as $\mathcal{D}$ increases. In addition, while the deterministic analysis of such systems becomes progressively more complex, the reduced system continues to provide a good approximation of the fixation probability and fixation time, as demonstrated in \fref{figMigrationSelectionBalance2}, inset. Finally, we would like to emphasise that we have here applied the reduction method to an extreme and very particular set of parameters, essentially testing the method to breaking point. This is done to demonstrate the quality of the approximation for large values of $s/N$.

\subsection{Hub}\label{secHub}

Having discussed the general predictions of the reduced model in both the neutral case and that in which selection is present, we now proceed to apply the results to a specific metapopulation topology, that of the hub or spoke (see \fref{hubFigure}). Our reasons for choosing such a system are twofold. Firstly the system possess symmetries which make it particularly suitable to an analytic treatment (though we stress that our method can also be used for more general systems). Secondly, such a structure allows us to investigate the behaviour of the model systematically as the number of demes increases.

Let us now consider the details of the system. We define the hub topology as one featuring a main deme which is connected to $\mathcal{D}-1$ satellite demes. The satellite demes themselves are entirely unconnected to one another. Migration probabilities along the connections are chosen so as to limit the parameter space but still allow for non-trivial behaviour. 


\begin{figure}[h!]
\centering
 \includegraphics[width=0.45\textwidth]{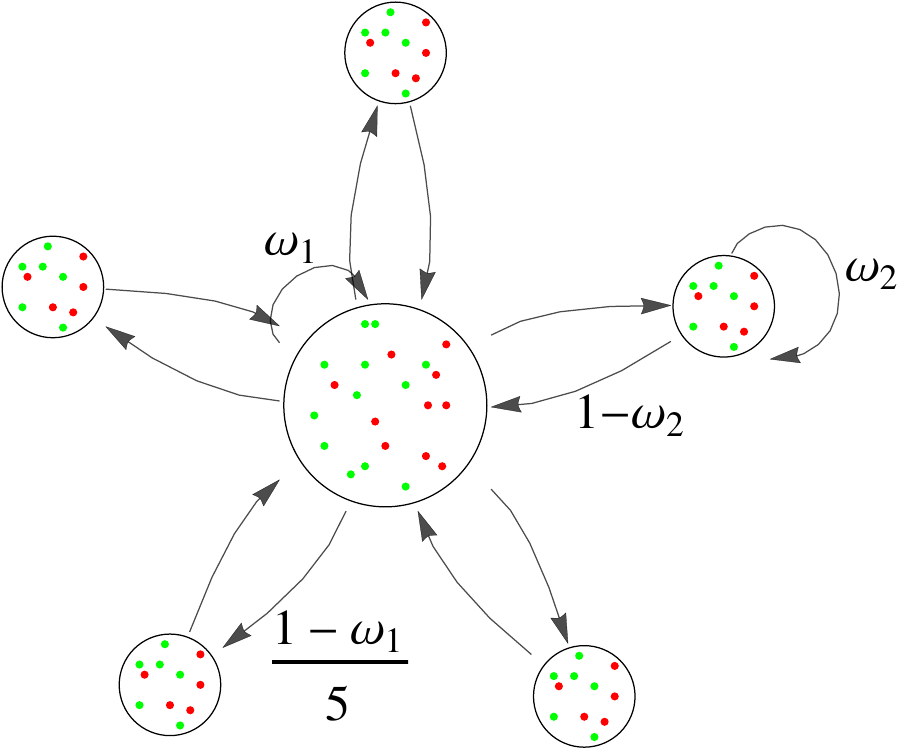}
\caption{Metapopulation model possessing a hub structure with $\mathcal{D}=6$. The central deme has a population of $\beta_{1}N$ while surrounding demes have a population of $\beta_{2}N$. Migration probabilities (conditional on origin island being first selected) in this case can be parametrised by the probability of remaining on a particular deme. The probability of remaining on the central deme is $\omega_{1}$, while the probability of migrating is dispersed equally over the satellite demes. The probability of remaining on a satellite deme is $\omega_{2}$, with probability $1 - \omega_{2}$ of migrating to the central deme.}
\label{hubFigure}
\end{figure} 


Recall the definition of the migration matrix $m$ in \Sref{secNeutralMigration}; previously we stated that the columns of $m$ were normalised such that the probability the offspring from a reproduction event would not migrate was equal to $1$ minus the total probability it would migrate, $m_{jj} = 1 - \sum_{i \neq j }^{\mathcal{D}}m_{ij}$. In this case however, since we have a more restricted geometry, we can instead parametrise the migration probabilities by the probability that the offspring \emph{remains} in the same deme as its parent. Defining $\omega_{1}$ as the probability that an offspring produced in the central deme does not migrate and $\omega_{2}$ the probability that an offspring from a satellite deme does not migrate, the normalised migration matrix for $\mathcal{D}$ demes is
\begin{eqnarray}\label{hubMigrationMatrix}
 m = \left(\begin{array}{ccccc}  \omega_{1}  & 1 - \omega_{2} & 1 - \omega_{2} & \ldots & 1 - \omega_{2}\\ 
                                    \frac{ 1 - \omega_{1} }{ (\mathcal{D}-1) } &  \omega_{2} & 0 & \ldots & 0 \\ 
                                  \frac{ 1 - \omega_{1} }{ (\mathcal{D}-1) }& 0 &  \omega_{2} & \ldots & 0 \\ 
                                  \vdots & \vdots & \vdots  & \ddots & \vdots\\ 
                                                \frac{ 1 - \omega_{1} }{ (\mathcal{D}-1) }& 0 & 0  &  0 &  \omega_{2}\end{array}\right) \, .
\end{eqnarray}

Further we take the central deme to have a population of $\beta_{1}N$, the satellite demes to have populations $\beta_{2}N$, and the birth rate in each deme be proportional to the island size, so that $f_{j}=\beta_{j}/\sum_{i=1}^{\mathcal{D}}\beta_{i}$.

We now apply the theory developed in \Sref{secReduction}. We begin by constructing the matrix $H$ from the migration matrix $m$, and island sizes $\bm{\beta}$.
Before proceeding further, we calculate the eigenvalues of $H$, as it is these that define the parameter range over which we would expect the approximation to work (see \sref{secReduction}). For convenience we introduce the quantities
\begin{eqnarray}
 \gamma_{1} &=& \beta_{1}^{2}(1 - \omega_{1}) \,, \\ \gamma_{2} &=& (\mathcal{D}-1)\beta_{2}^{2}(1 - \omega_{2}) \,, \\
 \gamma_{3} &=& \gamma_{1} + (\mathcal{D}-1)\gamma_{2} \, , \\\gamma_{4} &=& (\mathcal{D}-1) \beta_{1} \beta_{2} \left[ \beta_{1} + (\mathcal{D}-1) \beta_{2} \right]\, .
\end{eqnarray}
The first two eigenvalues are then given by
\begin{eqnarray}\label{hubEigenvalues}
 \lambda^{(1)} = 0 \, , \qquad
\lambda^{(2)} =  -\frac{ \gamma_{3} }{ \gamma_{4}} \,,
\end{eqnarray}
and for the remaining eigenvalues we find
\begin{eqnarray}
 \lambda^{(j)} = - \frac{\gamma_{1} }{\gamma_{4} }\, , \quad   j \geq 3 \, . \label{hubEigenvalue3}
\end{eqnarray}
By considering these eigenvalues we are already alerted to parameter regimes in which the reduced system could potentially give poor agreement with the full system. For instance, as we increase the number of demes in the system, $\mathcal{D}$, we find $\lambda^{(2)}$ tends to a finite quantity, $-(\omega_{2}-1)/\beta_{1}$. However, the remaining non-zero eigenvalues tend to zero with increasing deme number. One therefore must be cautious when applying the approximation technique to a hub system with a large number of satellite demes, as we expect that the approximation will break down if the magnitude of these eigenvalues approaches $N^{-1/2}$.

To obtain the reduced model in the neutral case, we need only calculate $\bm{u}^{(1)}$ (see \eref{defineBBar}). In the case where we look at second order effects in $s$, we must also calculate the remaining left- and right-eigenvectors. Since the system contains degenerate eigenvalues (\eref{hubEigenvalue3}) --- this is frequently the case in such highly symmetric systems --- the corresponding eigenvectors will not automatically be orthogonal. An orthogonal set must be constructed by taking linear combinations of these vectors, so that the orthonormality condition, \eref{orthonormal}, holds. The left-eigenvectors can be expressed as 
\begin{eqnarray}\label{hubLeft1}
 \bm{u}^{(1)} = \frac{1}{\gamma_{3}} 
		\left( \begin{array}{c} \gamma_{1} \\ \gamma_{2} \\ \vdots \\\gamma_{2} \end{array} \right) \, , \qquad
\bm{u}^{(2)} = \left( \begin{array}{c} - (\mathcal{D}-1) \\ 1 \\ \vdots \\ 1 \end{array} \right) \,,
\end{eqnarray}
and
\begin{eqnarray}\label{hubLeft2}
 u^{(j)}_{i} = \delta_{ij} - \frac{1}{j-2}\sum_{l=2}^{j-1} \delta_{li} \, , \quad  j \geq 3 \, .
\end{eqnarray}
The right eigenvectors meanwhile are given by 
\begin{eqnarray}\label{hubRight1}
 \bm{v}^{(1)} = \bm{1} \, , \qquad
 \bm{v}^{(2)} = \frac{1}{(\mathcal{D}-1)\gamma_{3} } \left( \begin{array}{c} -(\mathcal{D}-1) \gamma_{2}  \\ \gamma_{1} \\ \vdots \\ \gamma_{1}  \end{array}\right) \, , 
\end{eqnarray}
and
\begin{eqnarray}\label{hubRight2}
 v^{(j)}_{i} = \frac{j-2}{j-1} \left( \delta_{ij} - \frac{1}{j-2}\sum_{l=2}^{j-1} \delta_{li} \right) \, , \quad  j \geq 3 \, .
\end{eqnarray}
With these quantities in hand we can calculate the fixation probability and fixation time as defined in \Sref{secProbabilitiesNeutral} and \Sref{secProbabilitiesSelection}.

\subsubsection{Hub: s=0}\label{neutralHub}

In order to see how our reduced hub model compares against the full system with increasing  $\mathcal{D}$, we can look at how the fixation probability $Q(z_{0})$ and the fixation time $T(z_{0})$, normalised by the total system population, $N_{\rm{Tot}}$, changes with the initial condition $z_{0}$, while other parameters are kept fixed. The results are plotted in \fref{QTAsFuncOfDNeutral}, and we see that as $\mathcal{D}$ increases the approximation continues to provide good agreement with the exact Gillespie simulation of \eref{Meqn}, with transition rates given by \eref{transitionRatesNeutral1} and \eref{transitionRatesNeutral2}. 

As stated in \sref{secProbabilitiesNeutral}, the probability of fixation is only dependent on network structure through the projected initial condition, $z_{0}$. Since $z_{0}$ is held constant in this case, the probability of fixation does not change as the network structure is altered. We note that the behaviour of the fixation time as a function of the number of demes is non-trivial however. The results in \fref{QTAsFuncOfDNeutral} may be compared with the results for a single island of the same size as the total hub population; increasing the size of the island would give $r_{N}=1$ regardless of size. Here we see $r_{N}$ starts at a significantly higher value and decreases as the deme number (and hence the total population size) is increased.


\begin{figure}[h!]
\includegraphics[width=0.45\textwidth]{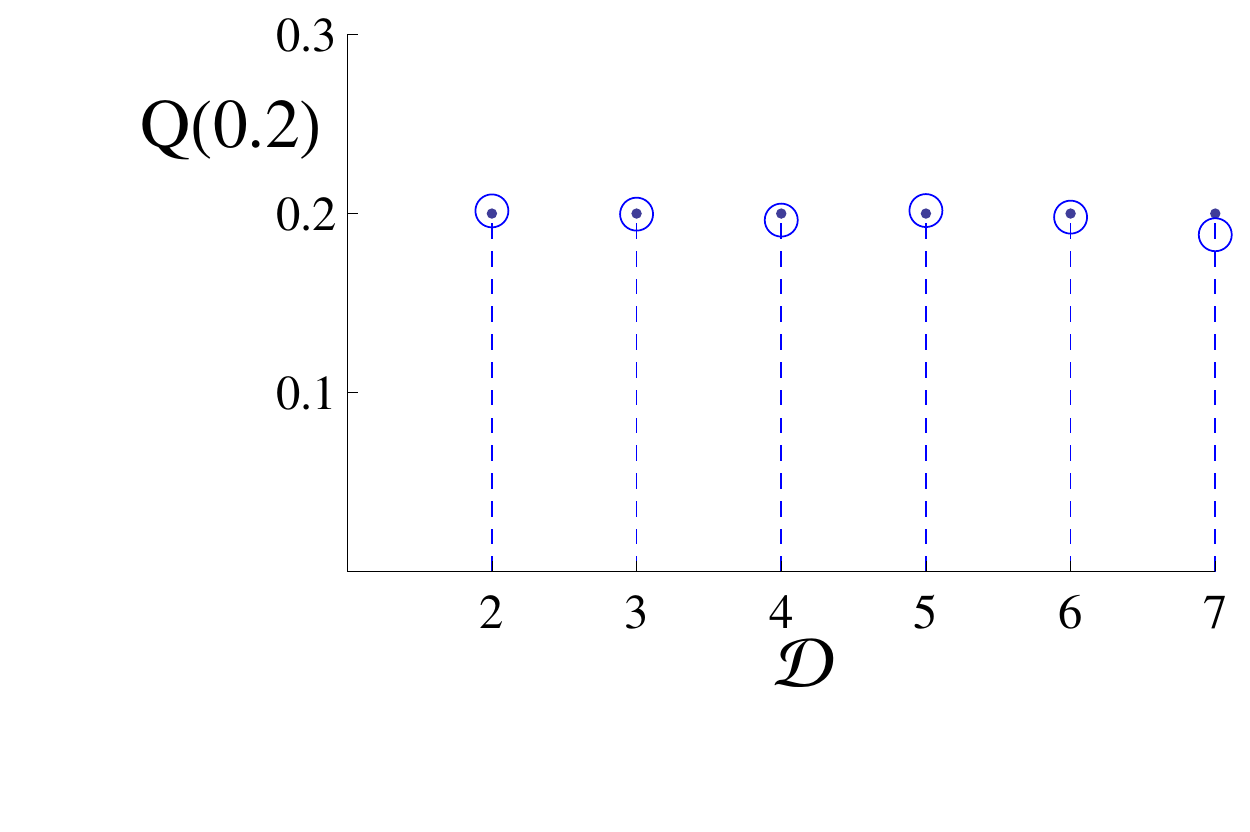}
\includegraphics[width=0.45\textwidth]{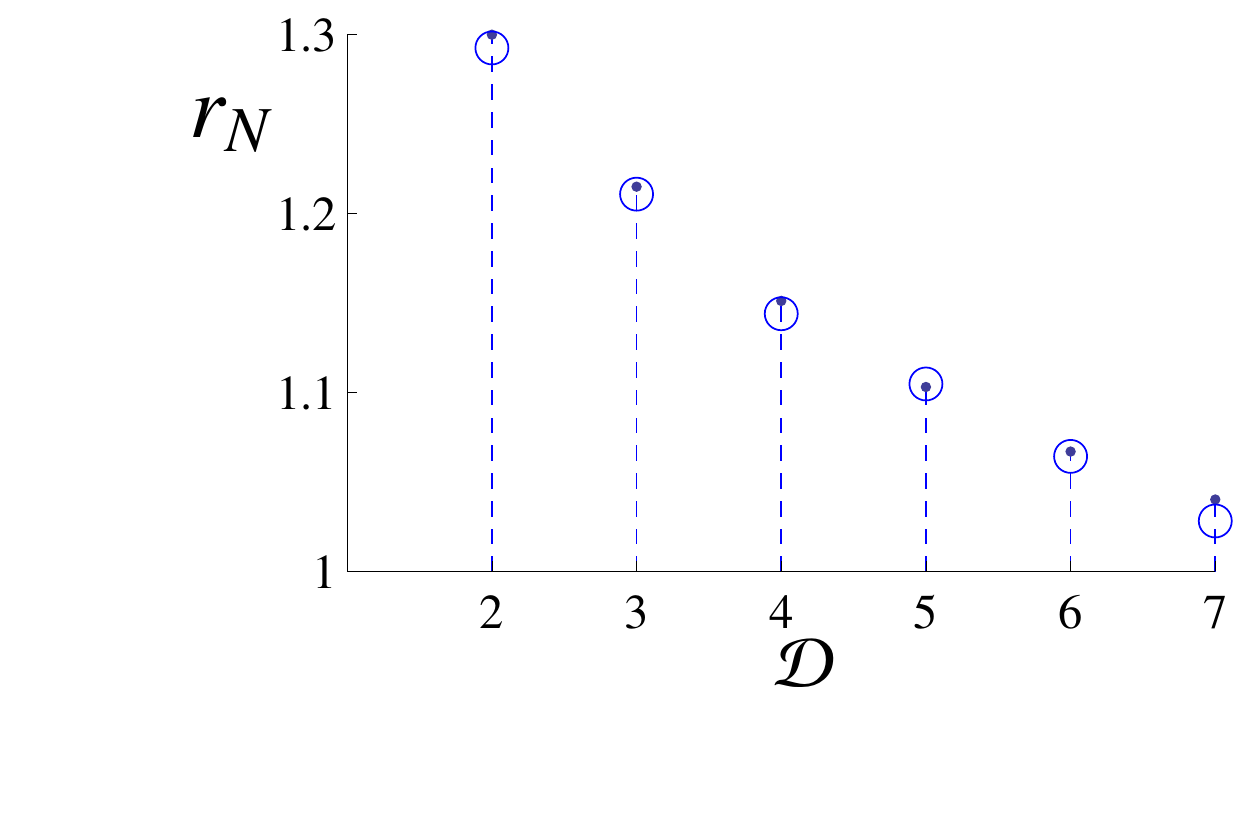}
\caption{Top panel: probability of fixation, $Q(z_{0})$, with fixed initial condition $z_{0}=0.2$, for the neutral hub model ($s=0$) plotted as a function of the number of demes, $\mathcal{D}$. Bottom panel: parameter $r_{N}$ as a function of $\mathcal{D}$ for the neutral hub model. Discrete analytic results are plotted as a dashed blue columns, and calculated from \eref{T_s_0} and \eref{defineRN}, and \eref{defineb1} calculated using \eref{hubLeft1}. Simulation results, plotted as circles, are the mean results from $6000$ runs. Parameters used in this example are $\omega_{1}=0.5$, $\omega_{2}=15/16$, $\beta_{1}=6$, $\beta_{2}=1$,  and finally $N=300$.}
\label{QTAsFuncOfDNeutral}
\end{figure}


\subsubsection{Hub with selection}\label{hubWithSelection}

Let us now incorporate selection into the general hub model described in \fref{hubFigure}. Suppose that the selection strength in the central deme, $1$, is moderated by $\alpha_{1}$ while the selection strength in the satellite demes is moderated by $\alpha_{2}$. The $\bar{A}(z)$ term for the system is given by \eref{ABar_s_N} and $\bar{B}(z)$ by \eref{defineBBar}. The parameters $a_{1}$, $a_{2}$ and $a_{3}$ are given by Eqs.~(\ref{definea1}), (\ref{definea2}) and (\ref{definea3}) which can now be easily calculated since we have the left- and right-eigenvectors of $H$ (equations \eref{hubLeft1}, \eref{hubLeft2} and \eref{hubRight1}, \eref{hubRight2} respectively). Their exact forms are too lengthy to be reproduced here, but are obtained by direct substitution. The results can then be tested against exact Gillespie simulations of the stochastic system defined by \eref{transitionRatesSelection}. As an example, let us compare two systems. 

In the first system we fix the number of demes to two with the first deme being defined as the central deme with population $\beta_{1}N$ and the second as the satellite deme with population $\beta_{2}N$. In deme one, the $A$ alleles experience a selective pressure $s\alpha_{1}$, while in the deme two, the satellite deme, the alleles experience a selective pressure $s\alpha_{2}$. 

In the second system, we again have a central deme with a population of $\beta_{1}N$, but we now fix the total population in each satellite deme to $N$ and vary the total deme number $\mathcal{D}$. Again the fitness in the central deme is equal to $s \alpha_{1}$ and the fitness in each satellite deme is equal to $s\alpha_{2}$. We can then say that in both systems, the number of individuals in the selective environments $s\alpha_{1}$ and $s\alpha_{2}$ are equal if $\beta_{2} = (\mathcal{D}-1)$. 

Na\"\i vely then, one might expect the systems to behave similarly for such metrics as fixation probability and fixation time, as $\beta_{2}$ and $\mathcal{D}-1$ respectively increase in each system. However, the analytical results predict very different behaviour. This is supported by simulation; in \fref{QTAsFuncOfDSelection}, a particular set of parameters is fixed and the size of the populations (moderated by $\beta_{2}$ in the first case and discretely by $\mathcal{D}$ in the second case) is increased. The probability of $A$ fixating decreases much more rapidly with an increasing number of satellite demes than the two deme system with increasing size of the second island. The time to fixation meanwhile increases more rapidly with the increasing number of satellite demes than that with two islands. 


\begin{figure}[h!]
\includegraphics[width=0.45\textwidth]{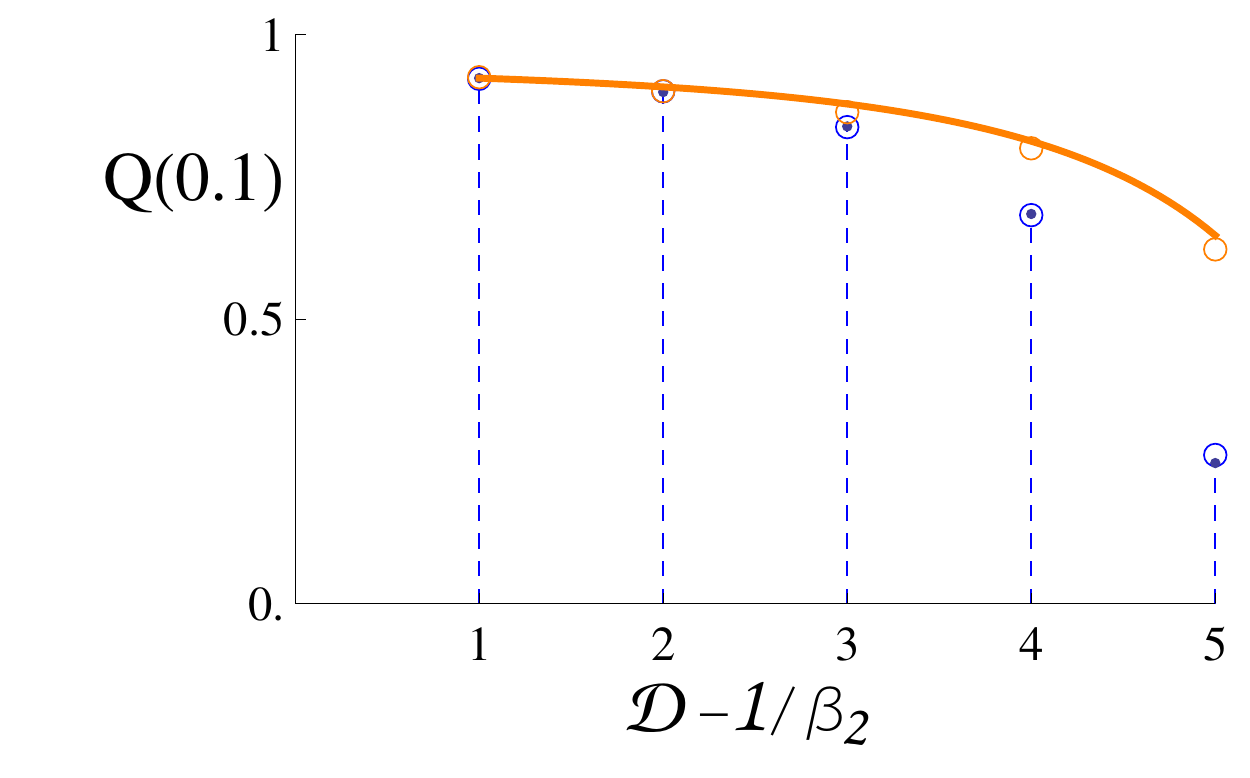}
\includegraphics[width=0.45\textwidth]{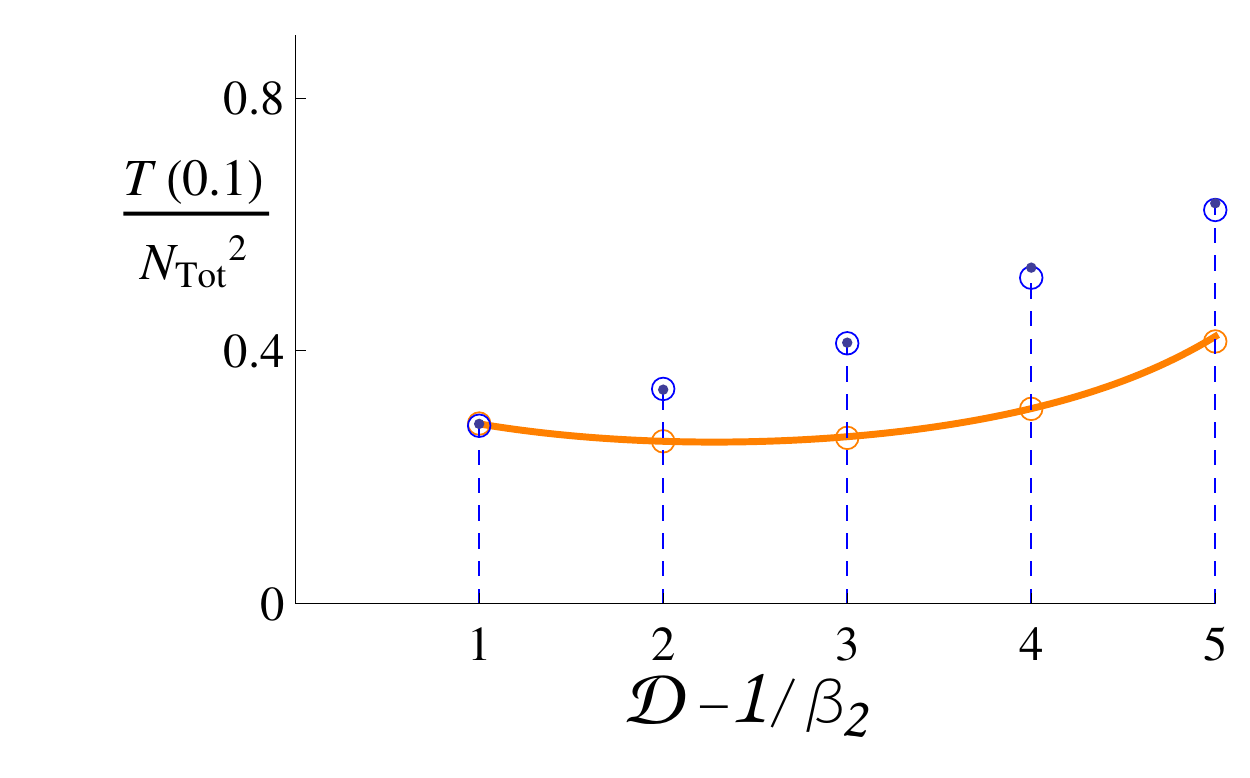}
\caption{Plots of fixation probability, $Q(z_{0})$ and fixation time, $T(z_{0})$ for two different models at $z_{0}=0.1$. The orange lines are obtained from the reduced model of a two-deme system in which $\beta_{2}$, the relative size of the second deme, is increased. The results from simulation of the system are shown as orange circles. The discrete values indicated by the blue dashed lines are obtained from a hub model in which the number of ($\mathcal{D}-1$) satellite demes is increased. Simulation results are plotted as blue circles. Simulation results are the mean of $2000$ runs. Parameters used are $s= 0.03$, $\alpha_{1}=1$, $\alpha_{2}=-1$, $\beta_{1}= 3$, $\beta_{2}= 1$, $\omega_{1}=0.625$, $\omega_{2}=0.9375$, and $N_{\rm Tot}= 300$.}
\label{QTAsFuncOfDSelection}
\end{figure}


\section{Conclusions}

Migration, along with mutation, selection and drift is one of the fundamental processes occurring in genetics, and island models were introduced very early on in the mathematical formulation of population genetics \cite{wright1931}. Yet theoretical studies of their general features are comparatively rare, and even the literature on specific models, while quite extensive, features no common approach.  Given the wide variety of methodologies and notations employed, it is particularly difficult to get an overview of theoretical work on island models. A major reason for this must be the perceived difficulty of formulating a stochastic version of the theory --- required if the population size is not assumed to be infinite --- and even more, the intractability of the resulting equations.

In this paper we have tried to address both of these problems. The first, that of formulating a stochastic version of the model, is the least difficult. We contend that the most systematic and straightforward way of proceeding is to first formulate the model at the `microscopic' level, that is, in terms of individuals which undergo the fundamental processes of birth, death, migration between islands, etc.. This can be achieved through the definition of states of the system (simply the numbers of individuals of different types on each island) and transition probabilities per unit time between these states. The latter are made up of combinatoric factors corresponding to sampling of specific types of individuals in the system, together with the rates at which the fundamental processes such as birth, death and migration occur. These transition probabilities per unit time, together with the usual assumption that the processes are Markovian, lead to a master equation (a Markov chain in continuous time) which governs the dynamics of the system.

This is a systematic procedure which results in an equation which completely specifies the stochastic dynamics, once initial conditions are given. However, the master equation is completely intractable in all situations of interest, and so starting from the earliest studies~\cite{fisher1922} a diffusion approximation has been made. Here the states are now the fraction of individuals of different types, which for large island sizes can be taken to be continuous variables to a good approximation. Within this approximation the master equation becomes a FPE which is equivalent to a set of SDEs. This is the starting point for many authors in their of studies of island models. However, we stress again that beginning from a master equation is both more natural and easier to interpret. It also avoids technical issues, such as trying to decide whether or not the process is of the It\={o} or Stratonovich type \cite{gardiner2009}. This procedure was described in \cite{mckaneModels2007} for the neutral case and islands all having the same number of individuals. Here we have extended it to include the effects of selection and differing island size.

The equations resulting from this procedure are, however, still formidably complicated. The FPEs are multidimensional partial differential equations and the set of SDEs are coupled, nonlinear and have multiplicative noise. The second, and more substantive, aim of this paper has been to show that the application of a procedure which we recently devised~\cite{projectionPhys}, allows the FPE to be reduced to a much simpler FPE, or equivalently allows the set of SDEs to be reduced to a single SDE. The resulting equations are then amenable to analysis. The technique is simple to understand, being based on the elimination of modes in the system which decay quickly, and gives rise to explicit formulae for the parameters in the final, simplified equations. These parameters can be straightforwardly calculated from the network structure of the islands. 

We have applied this method to systems defined by a range of different networks, comparing the results we obtain to a direct simulation of the IBM. The method gives excellent results for most networks and parameter values. Where it does not work so well, there are reasons why we would expect this. For instance, we require the magnitude of the real part of the non-zero eigenvalues of $H$ to be greater than both $s$ and $N^{-1/2}$, so that the separation of timescales is sufficient to apply the approximation. Furthermore, we assume that parameters modelling similar effects in the model are not of completely different orders. Therefore, no island is assumed to be an order or two of magnitude bigger than other islands (all $\beta_i$ are of order one). Even if these conditions are violated, the approximation may work reasonably well. We expect that the elimination of fast variables will still be possible in such cases, so a different calibration could describe some of these situations. For example, the diagonal elements of $H$ could be made to scale with $N^{-1/2}$, which would result in a different set of formulae. In \cite{mckaneUtterance}, the diagonal elements were chosen to scale like $N^{-1}$, which accounts in part for the differences between the results given in that paper and those given here.

We have chosen to work only to order $s^2$, although the technique is capable of being generalised to higher orders in $s$. Given the typical size of selection strengths, working to this order is entirely reasonable, and indeed most authors only keep terms of order $s$. We should emphasise that although we frequently compare $s$ numerically to $N^{-1/2}$ or $N^{-1}$, we do not, unlike some authors, set $s$ equal to $N^{-1/2}$ or $N^{-1}$; $s$ and $N$ are two independent parameters. One of the reasons we go to order $s^2$ is to show that the analysis of the stochastic aspects of migration-selection balance could be considerably extended using our results. 

The model we present here is slightly different from the previous work we have addressed which concerns migration-selection balance. However, we have shown that our method, being more general, can be used to investigate a broader range of parameters than previously attempted, including non-symmetric migration, arbitrary deme topology and and arbitrary range of selective pressures for each deme. Further, we note that the mechanism that allows our approximation to work so successfully is the dominance of the large and linear effect of migration (in our model embodied by the matrix $H$) over smaller non-linear and stochastic terms.  

Conceivably then, in addition to the points we have mentioned, there may be other ways of extending and generalising our treatment. For instance, we have imposed fixed population size for each island. This is traditional in the context of population genetics, and the Moran process has this assumption at its heart. However it should be possible to relax this assumption. Other processes, such as mutation, could also be included. We are in the process of investigating some of the questions, and hope to report on them in the future.

The formulation and subsequent analysis of stochastic effects or of migration in population genetics is often perceived as being difficult, and systems where both are important doubly so. We believe this need not be the case. The formulation can be made systematic and intuitively appealing by starting from an IBM and invoking the diffusion approximation. The analysis of the resulting equations is possible through the elimination of the fast variables, leading to much simpler equations which are amenable to analysis. We hope that this methodology expounded here will be taken up by other researchers and will lead to the analysis of more complex and realistic models.

\section*{Acknowledgement}
We thank Diana Garcia L\'{o}pez for useful discussions. G.W.A.C. thanks the Faculty of Engineering and Physical Sciences, University of Manchester for funding through a Dean's Scholarship.

\begin{appendix}

\section{The transition rates and the derivation of the Fokker-Planck equation}\label{appKMExpansion}

Here we describe the mathematical machinery employed in moving from a description of the system in terms of the probability transition rates (e.g. Eqs.~(\ref{transitionRatesNeutral1}) and (\ref{transitionRatesNeutral2})), to a description in terms of the Fokker-Planck equation or the equivalent stochastic differential equations. 

The state of the system at any moment in time is given the vector $\bm{n}=(n_1,\ldots n_\mathcal{D})$, where $n_i$ specifies the number of individuals carrying allele $A$ on island $i$. The transition rates $T(\bm{n}|\bm{n}')$, are measures of the probability per unit time that the system moves from state $\bm{n}'$ to $\bm{n}$. The transition rates for the neutral migration model are given by  Eqs.~(\ref{transitionRatesNeutral1}) and (\ref{transitionRatesNeutral2}), where they are derived from a consideration of some standard combinatorics. Here we shall discuss the derivation of the model including selection (see \Sref{secSelectionMigration}), which reduces to the neutral model as the selection strength $s$ tends to zero.

Suppose that now, rather than individuals carrying allele $A$ or $B$ being just as likely to reproduce, that one has a fitness advantage over the other. In an analogous manner to the one deme case of \Sref{secSelectionReview}, we introduce weighting vectors $\bm{w}_{A}$ and $\bm{w}_{B}$ which describe the relative likelihood of type $A$ or $B$ reproducing on any of the islands. The probability per unit time of a type $A$ individual on island $i$ reproducing is then given by the number of $A$ individuals on island $i$, $n_{i}$, multiplied by the relative fitness of allele $A$ on island $i$, $[\bm{w}_{A}]_{i}$, normalised by the total fitness of the population of island $i$, $n_{i}[\bm{w}_{A}]_{i} + (N-n)[\bm{w}_{B}]_{i}$. The progeny may then either remain in its own deme or migrate to another, based on the migration rate matrix $G$. The individual it displaces is chosen neutrally, based on local allele frequencies. The probability transition rates are then given by
\begin{eqnarray*}
&& T(n_{i}+1|  n_{i} ) = \\ & &\sum_{j=1}^{\mathcal{D}}  \frac{(\beta_i N-n_{i})}{\beta_{i}N - \delta_{ij}}G_{ij} \frac{ [\bm{w}_{A}]_{j}n_{j}}{[\bm{w}_{A}]_{j}n_{j}+  [\bm{w}_{B}]_{j}(\beta_{j}N-n_{j})}, \\
&& T(n_{i}-1| n_{i}) = \\ & &  \sum_{j=1}^{\mathcal{D}}   \frac{n_{i}}{\beta_{i}N-\delta_{ij}}G_{ij}\frac{ [\bm{w}_{B}]_{j}(\beta_{j}N- n_{j})}{[\bm{w}_{A}]_{j}n_{j}+  [\bm{w}_{B}]_{j}(\beta_{j}N-n_{j})},
\end{eqnarray*}
using analogous arguments to those used to obtain the neutral transition rates. 

We can further simplify these expressions by setting $[\bm{w}_{B}]_{i}=1$ and $[\bm{w}_{A}]_{i} = 1 + s \alpha_{i}$ for each island. The parameter $s$ is an indicative selection strength, while the elements of $\bm{\alpha}$ will be assumed to be of order $1$ and will primarily be used to signify the direction of selection. If $\alpha_{i}>0$ then $[\bm{w}_{A}]_{i}>[\bm{w}_{B}]_{i}$ and allele $A$ is advantageous on island $i$, while if $\alpha_{i}<0$, allele $A$ will be deleterious on that island. Finally, if we assume that the selection strength $s$ is small, we can express the above transition rates as a Taylor series in $s$. Suppressing the dependence of $T(\bm{n}|\bm{n'})$ on states that do not vary in a particular transition, we obtain
\begin{eqnarray*}
 T( n_{i}+1 |  n_{i}) = & &  \sum_{j= 1}^{\mathcal{D}}  \frac{(\beta_i N-n_{i})}{\beta_{i}N - \delta_{ij}}G_{ij} \times \\
                               & &\left( \frac{n_{j}}{\beta_{j}N} + s \alpha_{j} \frac{n_{j}(\beta_{j}N-n_{j})}{(\beta_{j}N)^{2}} \right. \nonumber \\ &&\left. - s^{2}\alpha_{j}^{2} \frac{n_{j}^{2}(\beta_{j}N - n_{j})}{(\beta_{j}N)^{3}} + \mathcal{O}(s^{3}) \right) \, , \\
 T(n_{i}-1 | n_{i} ) =  & & \sum_{j=1}^{\mathcal{D}}   \frac{n_{i}}{\beta_{i}N-\delta_{ij}}G_{ij}\times \\
                             & & \left( 1 - \frac{n_{j}}{\beta_{j}N} - s \alpha_{j} \frac{n_{j}(\beta_{j}N-n_{j})}{(\beta_{j}N)^{2}} \right. \nonumber \\ &&\left. + s^{2}\alpha_{j}^{2} \frac{n_{j}^{2}(\beta_{j}N - n_{j})}{(\beta_{j}N)^{3}} + \mathcal{O}(s^{3}) \right) \, .
\end{eqnarray*}

The dynamics of the system can be described by a master equation, as explained 
\Sref{secNeutralReview} in the case of a single island without selection. It is a simple generalisation of Eq.~(\ref{masterEquation1D}) with now the state being specified by the vector $\bm{n}$:
\begin{eqnarray}
  \frac{d p(\bm{n},t)}{d t} &=& \sum_{i=1}^{\mathcal{D}} \left[ T(n_{i}|n_{i}-1)p(n_{i}-1,t) \right. \nonumber \\ && \left. - T(n_{i}+1|n_{i})p(n_{i},t) \right] \nonumber \\
			    &+&  \sum_{i=1}^{\mathcal{D}} \left[ T(n_{i}|n_{i}+1)p(n_{i}+1,t) \right. \nonumber \\ && \left. - T(n_{i}-1|n_{i})p(n_{i},t) \right]\,.
  \label{Meqn}
\end{eqnarray}
It is to this master equation, with the transition rates given above, that we wish to apply the Kramers-Moyal expansion to obtain the Fokker-Planck equation.

The dynamics can be seen to be that of a one-step process; any one transition can only move the system from an initial state $\bm{n}'= (n_{1}, \ldots n_{i}, \ldots  n_{\mathcal{D}})$ to the adjacent states $\bm{n}'= (n_{1}, \ldots n_{i}\pm 1, \ldots  n_{\mathcal{D}})$. We can exploit this fact notationally; introducing new state variables $\bm{x}$ such that $x_{i} = n_{i}/\beta_{i}N$, we can write $f^{+}_{i}(x_i)$ and $f^{-}_{i}(x_i)$ as shorthand for the transition rates (in terms of the new variables) for moving up to state $x_{i}+1/\beta_{i}N$ or down in state $x_{i}-1/\beta_{i}N$ from initial state $\bm{x}'$. This gives
\begin{eqnarray}
& & f^{+}_{i}(x_{i}) = \frac{ G_{ii}(1 - x_{i}) }{1 - (\beta_{i}N)^{-1}} \times \nonumber \\ && \left[x_{i} + s\alpha_{i}x_{i}(1-x_{i})  - s^{2}\alpha_{i}^{2}x_{i}^{2}(1-x_{i}) \right] \nonumber\\
& & + (1-x_{i})\sum_{j\neq i}^{\mathcal{D}}G_{ij}\left[x_{j} + s\alpha_{j}x_{j}(1-x_{j}) \right. \nonumber \\ && \hspace{2cm} \left. - s^{2}\alpha_{j}^{2}x_{j}^{2}(1-x_{j}) \right] + \mathcal{O}(s^{3})\, , \nonumber \\
& & f^{-}_{i}(x_{i}) = \frac{ G_{ii}x_{i} }{1 - (\beta_{i}N)^{-1}} \times \nonumber \\  &&       \left[ (1 -x_{i}) - s\alpha_{i}x_{i}(1-x_{i}) + s^{2}\alpha_{i}^{2}x_{i}^{2}(1-x_{i}) \right] \nonumber \\
& & + x_{i}\sum_{j\neq i}^{\mathcal{D}}G_{ij}\left[ (1 -x_{j}) - s\alpha_{j}x_{j}(1-x_{j})  \right. \nonumber \\ && \hspace{2cm} \left. + s^{2}\alpha_{j}^{2}x_{j}^{2}(1-x_{j}) \right] + \mathcal{O}(s^{3}) \, .
\label{fs}
\end{eqnarray}
For now let us leave the specific form of these transition rate functions alone, pausing only to note that the typical deme size, $N$, now only appears in the first term of $f^{+}_{i}(x_{i})$ and $f^{-}_{i}(x_{i})$. 

We now re-express the master equation in terms of the transition rates $f^{+}_{i}(x_{i})$ and $f^{-}_{i}(x_{i})$:
\begin{eqnarray}
\frac{dp}{dt} &=& \sum_{i=1}^{\mathcal{D}} \left[ f^{+}_{i}(x_{i}- \frac{1}{\beta_{i}N})p(x_{i}-\frac{1}{\beta_{i}N},t) \right. \nonumber \\ && \hspace{2cm} \left.\vphantom{\frac{1}{\beta_{i}N})}  - f^{+}_{i}(x_{i})p(x_{i},t) \right] \nonumber \\
&+&  \sum_{i=1}^{\mathcal{D}} \left[ f^{-}(x_{i}+\frac{1}{\beta_{i}N})p(x_{i}+\frac{1}{\beta_{i}N},t) \right. \nonumber \\ && \hspace{2cm} \left. \vphantom{\frac{1}{\beta_{i}N})} - f^{-}_{i}(x_{i})p(x_{i},t) \right]\,.
\label{master_f}
\end{eqnarray}
Assuming the typical deme population $N$ to be large, we can carry out a Taylor expansion in $N^{-1}$; this is in effect the Kramers-Moyal expansion~\cite{gardiner2009,risken1989}. The right-hand side of the master equation (\ref{master_f}) becomes
\begin{eqnarray*}
&-& \sum_{i=1}^{\mathcal{D}} \left\{ \left(\frac{1}{\beta_{i}N}\right) \frac{\partial}{\partial x_{i}} \left[f^{+}_{i}(x_{i})p(x_{i},t) \right]   \right\} \nonumber \\
& & + \frac{1}{2!}\sum_{i=1}^{\mathcal{D}} \left\{ \left(\frac{1}{\beta_{i}N}\right)^{2} \frac{\partial^{2}}{\partial x_{i}^{2}} \left[ f^{+}_{i}(x_{i})p(x_{i},t) \right] \right\} \\
&+& \sum_{i=1}^{\mathcal{D}} \left\{ \left(\frac{1}{\beta_{i}N}\right) \frac{\partial}{\partial x_{i}} \left[f^{-}_{i}(x_{i})p(x_{i},t) \right]   \right\} \nonumber \\
& & + \frac{1}{2!}\sum_{i=1}^{\mathcal{D}} \left\{ \left(\frac{1}{\beta_{i}N}\right)^{2} \frac{\partial^{2}}{\partial x_{i}^{2}} \left[ f^{-}_{i}(x_{i})p(x_{i},t) \right] \right\},
\end{eqnarray*}
plus terms in $N^{-3}$ and higher.

We now return to the terms in $f^{+}_{i}$ and $f^{-}_{i}$ which involve $N$. They are identical to lowest order in $s$, and equal
\begin{equation}
\frac{ G_{ii}(1 - x_{i})x_{i} }{1 - (\beta_{i}N)^{-1}} \,.
\label{extra_contrib}
\end{equation}
Now in the master equation $f^{+}_{i}$ and $f^{-}_{i}$ appear with different signs in the terms involving the first derivative, and so they cancel. Although their contributions add in the terms involving the second derivative, if we expand the expression (\ref{extra_contrib}) in powers of $N^{-1}$ we see that these give $\mathcal{O}(N^{-3})$ contributions in the Kramers-Moyal expansion, which we are discarding. By the same argument, the terms in $f^{+}_{i}$ and $f^{-}_{i}$ which involve $N$ and powers of $s$ will also give $\mathcal{O}(N^{-3})$ contributions when multiplying the second derivative, and so can also be discarded. Finally, when these $s$-dependent terms multiply the first derivative, they will give contributions $s/N^2$ and $s^2/N^2$, but we will not include such terms in the diffusion matrix $B$ (see below), and so we do not include them in this context either. So, in summary, the $N$ dependence which appears in $f^{+}_{i}$ and $f^{-}_{i}$ in Eq.~(\ref{fs}) may be omitted to the order we are working, and the only $N$ dependence is that shown explicitly in the Kramers-Moyal expansion of the right-hand side of the master equation.

We now define 
\begin{eqnarray}
A_{i}(\bm{x}) &=& \frac{1}{\beta_i}\,\left[ f^{+}_{i}(\bm{x}) 
- f^{-}_{i}(\bm{x}) \right], \nonumber \\
B_{ii}(\bm{x}) &=& \frac{1}{\beta^2_i}\,\left[ f^{+}_{i}(\bm{x}) 
+ f^{-}_{i}(\bm{x}) \right].
\label{AandB}
\end{eqnarray}
With these definitions the expansion of the master equation in inverse powers of $N$ takes the form
\begin{eqnarray}
 \frac{\partial p(\bm{x},t)}{\partial t} = &-& \frac{1}{N} \sum_{i=1}^{\mathcal{D}} \frac{\partial}{\partial x_{i}} \left[A_{i}(\bm{x})p(\bm{x},t)\right] \nonumber \\ &+& \frac{1}{2N^{2}}\sum_{i=1}^{\mathcal{D}}\frac{\partial^{2}}{\partial x_{i}^{2}} \left[B_{ii}(\bm{x})p(\bm{x},t)\right].
\end{eqnarray}
Substituting the explicit forms for $f^{\pm}_{i}$ given by Eq.~(\ref{fs}) into Eq.~(\ref{AandB}) gives the elements of the vector $\bm{A}(\bm{x})$ as 
\begin{eqnarray*}
 A_{i}(\bm{x}) =&& \frac{1}{\beta_{i}}\left \{ \sum_{j\neq i}^{\mathcal{D}} G_{ij}(x_{j} - x_{i}) \right. \nonumber \\ 
      & &+ s \sum_{j=1}^{\mathcal{D}} G_{ij}\alpha_{j}x_{j}(1-x_{j})   \nonumber \\
 && \left. - s^{2} \sum_{j=1}^{\mathcal{D}} G_{ij}\alpha_{j}^{2}x_{j}^{2}(1-x_{j}) \right \} + \mathcal{O}(s^{3}),
\end{eqnarray*}
and a diagonal diffusion matrix with elements given by
\begin{eqnarray}
 B_{ii}(\bm{x}) =  && \frac{1}{\beta_{i}^{2}} \left\{  x_{i}\sum_{j=1}^{\mathcal{D}}  G_{ij}  +\sum_{j=1}^{\mathcal{D}}  G_{ij}x_{j}   \right. \nonumber \\
  && \hspace{1cm} \left.- 2 x_{i}\sum_{j=1}^{\mathcal{D}}  G_{ij}  x_{j}        \right\} +  \mathcal{O}(s) \, .
\end{eqnarray}
The truncation of the series in $s$, should be chosen to be consistent with the truncation in the expansion in terms of $N$. This will clearly depend on the assumed size of $s$. If one sets $s=0$, the above model reduces to that stated for the neutral case, \eref{neutralDrift} and \eref{neutralDiff}. 
Further, it can be shown that the Fokker-Planck equation is equivalent~\cite{gardiner2009,risken1989} to the SDE
\begin{equation}
 \frac{dx_i}{d\tau} = A_{i}(\bm{x}) + \frac{1}{\sqrt{N}}\eta_i(\tau),
\end{equation}
defined in the sense of It\={o}~\cite{vankampen2007}. Here $\tau = t/N $, $\bm{\eta}(\tau)$ is a Gaussian white noise term such that $\langle \eta_i(\tau) \rangle = 0$ and $\langle \eta_{i}(\tau)\eta_{j}(\tau') \rangle = B_{ij}(\bm{x})\,\delta(\tau-\tau')$, although for the present model $B_{ij}(\bm{x})=0$ if $i \neq j$.

\section{Solution to the equation for the probability of fixation}\label{appSol}

The probability of fixation in the reduced system, $Q(z_0)$, is found from the backward equation corresponding to the FPE (\ref{generalFPE1D}), in exactly the same way that the equation for the probability of fixation in the single island case (\ref{BFPQ}), is found from the backward equation corresponding to Eq.~(\ref{FPE1D}). Therefore the equation reads
\begin{equation}
\frac{\bar{A}(z_0)}{N}\,\frac{dQ}{dz_0} + 
\frac{\bar{B}(z_0)}{2N^2}\,\frac{d^2Q}{dz^2_0}  = 0,
\label{BFPQ_z}
\end{equation}
where $z_0$ is the initial starting point on the centre manifold (or slow subspace). The boundary conditions are as for the single island case, that is, $Q(0)=0$ and $Q(1)=1$. In this appendix we discuss the analytic solution of Eq.~(\ref{BFPQ_z}) when $\bar{A}(z_0)$ and $\bar{B}(z_0)$ are given by \eref{ABar_s_N} and \eref{defineBBar}.

The result for the neutral case and to linear order in $s$ have the same form as in the one-island case, and are well known~\cite{ewens2004}. When $s=0$, $\bar{A}(z_0)=0$, and so the solution of Eq.~(\ref{BFPQ_z}) subject to the boundary conditions is simply $Q(z_0)=z_0$. At linear order in $s$, $\bar{A}(z)=sa_{1}z(1-z)$, and a straightforward integration of Eq.~(\ref{BFPQ_z}) gives Eq.~(\ref{1D_Q_s_N}), albeit with extra factors of $a_1$ and $b_1$ and with $x_0$ replaced by $z_0$ (see Eqs.~(\ref{Q_s_N}) and (\ref{Q_s_N_Asymp})). 

To second order in $s$, $\bar{A}(z)$ may be written in the form (\ref{ABar_s_N_compact}), while $\bar{B}(z)$ is still given by Eq.~(\ref{defineBBar}). The equation for the probability of fixation (\ref{BFPQ_z}) now takes the form 
\begin{equation*}
\frac{s}{N} z_{0} (1-z_{0})(k_{1} - s k_{2} z_{0})\frac{d Q}{d z_{0}}   + \frac{1}{N^{2}} b_{1} z_{0}(1-z_{0}) \frac{d^{2}Q}{dz_{0}^{2}} = 0.
\end{equation*}
Integrating with respect to $z_{0}$ we arrive at the equation
\begin{eqnarray*}
\frac{d Q}{d z_{0}} = c_{1} \exp{ \left[ -\frac{Ns}{b_{1}}(k_{1}z_{0}-\frac{sk_{2}}{2}z_{0}^{2})  \right] },
\end{eqnarray*}
where $c_{1}$ is a constant of integration yet to be determined and where we note from Eq.~(\ref{defineb1}) that $b_{1}>0$. 

If $k_2 = 0$, the calculation is identical to that carried out to first order in $s$, \eref{Q_s_N_Asymp}, but with $a_1$ replaced by $k_1$. If $k_{2}\neq 0$, we may complete the square in the exponent to find
\begin{eqnarray*}
\frac{d Q}{d z_{0}} = c_{1} \exp{ \left[-\frac{N k_{1}^{2}}{2 b_{1} k_{2}} \right]} \exp{ \left [\frac{N}{2 b_{1} k_{2}} (s k_{2} z_{0} - k_{1})^{2} \right] }.
\end{eqnarray*}
We now change variables from $z_0$ to $l$, where
\begin{eqnarray}\label{defineL}
 l = \sqrt{\frac{N}{2 b_{1} |k_{2}|}}(s k_{2} z_{0} - k_{1}),
\end{eqnarray}
to obtain
\begin{equation}
\label{dQ_by_dl}
\frac{dQ}{dl} = \left\{ \begin{array}{ll} 
- c_2 \exp{ ( - l^2 ) }, & \mbox{\ if $k_2 < 0$} \\ \\
c_2 \exp{ ( l^2 ) }, & \mbox{\ if $k_2 > 0$},
\end{array} \right.
\end{equation}
where
\begin{equation}
c_2 = \frac{c_{1}}{s} \sqrt{\frac{2b_{1}}{|k_{2}|N}} 
\exp\left\{ -\frac{N k_{1}^{2}}{2b_{1} k_{2}} \right\},
\end{equation}
is another constant.

The integrals over the exponentials in Eq.~(\ref{dQ_by_dl}) can be carried out in terms of functions related to the error function, namely the complementary error function~\cite{handbook1965}
\begin{equation}
\mathrm{erfc} y = 1 - \mathrm{erf} y = 1 - \frac{2}{\sqrt{\pi}} 
\int^{y}_{0}\,e^{-l^{2}}\,dl,
\label{erfc}
\end{equation}
and the imaginary error function~\cite{HTFS}
\begin{equation}
\mathrm{erfi} y = \frac{2}{\sqrt{\pi}} \int^{y}_{0}\,e^{l^{2}}\,dl.
\label{erfi}
\end{equation}
Implementing the boundary conditions $Q(l(z_{0}=0))=0$ and $Q(l(z_{0}=1))=1$, one finds 
\begin{equation}
Q(z_0) = \frac{ 1 - \chi(z_0)}{1 - \chi(1)},
\label{general_Q}
\end{equation}
where 
\begin{equation}
\label{defintion_chi_negative}
\chi(z_0) = \frac{\mathrm{erfc}(l(z_0))}{\mathrm{erfc}(l(0))}, \ \ 
\mathrm{if\ } k_2 < 0,
\end{equation}
and
\begin{equation}
\label{defintion_chi_positive}
\chi(z_0) = \frac{\mathrm{erfi}(l(z_0))}{\mathrm{erfi}(l(0))}, \ \ 
\mathrm{if\ } k_2 > 0.
\end{equation}

If $l$ is large, then asymptotic forms can be used to simplify both the complementary error function and the imaginary error function \cite{handbook1965,HTFS}:
\begin{equation}
\mathrm{erfc}(l) = \frac{e^{-l^{2}}}{\sqrt{\pi}l} \left[ 1 + \mathcal{O}\left( 
\frac{1}{\l^2} \right) \right],
\label{asymp_erfc}
\end{equation}
and
\begin{equation}
\mathrm{erfi}(l) = \frac{e^{l^{2}}}{\sqrt{\pi}l} \left[ 1 + \mathcal{O}\left( 
\frac{1}{\l^2} \right) \right].
\label{asymp_erfi}
\end{equation}

\section{Calculation of the mean time to fixation}\label{appMTF}

The mean time to fixation in the reduced system, $T(z_0)$, is found from the backward equation corresponding to the FPE (\ref{generalFPE1D}), in exactly the same way that the equation for mean time to fixation in the single island case (\ref{BFPT}), is found from the backward equation corresponding to Eq.~(\ref{FPE1D}). Therefore the equation reads
\begin{equation}
\frac{\bar{A}(z_0)}{N}\,\frac{dT}{dz_0} + 
\frac{\bar{B}(z_0)}{2N^2}\,\frac{d^2T}{dz^2_0}  = -1,
\label{BFPT_z}
\end{equation}
where $z_0$ is the initial starting point on the centre manifold (or slow subspace). The boundary conditions are as for the single island case, that is, $T(0)=0$ and $T(1)=0$. In this appendix we discuss the analytic solution of Eq.~(\ref{BFPT_z}) when $\bar{A}(z_0)$ and $\bar{B}(z_0)$ are given by \eref{ABar_s_N} and \eref{defineBBar}.

The result for the neutral case is well known~\cite{ewens2004}. Setting $s=0$ in \eref{ABar_s_N} gives $\bar{A}(z_0)=0$, and direct integration of Eq.~(\ref{BFPT_z}) gives Eq.~(\ref{1D_T_s_0}), albeit divided by a factor of $b_1$ and with $x_0$ replaced by $z_0$. At order $s$, $\bar{A}(z)=s a_1 z(1-z)$, and so the equation for $T(z_0)$ becomes
\begin{equation}
\frac{\sigma z_{0}(1-z_{0})}{M}\,\frac{dT}{dz_0} + 
\frac{z_{0}(1-z_{0})}{M^2}\,\frac{d^2T}{dz^2_0}  = -1,
\label{BFPT_z_1}
\end{equation}
where we have defined new parameters $M=N/\sqrt{b_{1}}$ and $\sigma = a_{1}s/\sqrt{b_{1}}$. The reason for introducing these new parameters, other than on grounds of simplicity, is that Eq.~(\ref{BFPT_z_1}) is exactly the equation found in the single island case with selection.

To solve it we introduce $\phi(z_0)=dT/dz_{0}$, so that the equation now reads
\begin{equation}
\frac{d\phi}{dz_0} + M\sigma \phi = - \frac{M^2}{z_{0}(1-z_{0})}.
\label{phi_eqn}
\end{equation}
This equation is difficult to deal with analytically and numerically because of the singularities on the right-hand side at precisely the values of $z_0$ where we need to impose the boundary conditions. One can avoid this problem by writing $\phi =\phi_{0} + \phi_{s}$, and choosing $\phi_0$ so that the term $d\phi_0/dz_0$ cancels the right-hand side of Eq.~(\ref{phi_eqn}). This choice means that $\phi_0$ is simply the $s=0$ solution, and the equation for $\phi_s$ is then
\begin{equation}
\frac{d\phi_s}{dz_0} + M\sigma \phi_s = - M\sigma\phi_0 = M^{3}\sigma \left[ \ln z_{0} - \ln (1-z_{0}) \right],
\label{phi_eqn_1}
\end{equation}
which on the left-hand side is exactly the same as the equation for $\phi$, but with a right-hand side which is less divergent as $z_{0} \to 0$ or $z_{0} \to 1$. Although this right-hand side is still divergent, its integral is not, which is all that we need. If we do require a convergent expression we can repeat the process, and write $\phi_{s}=\phi_{1} + \phi_{2}$, choosing $\phi_1$ so that the term $d\phi_1/dz_0$ cancels the right-hand side of Eq.~(\ref{phi_eqn_1}). 

We can now multiply Eq.~(\ref{phi_eqn_1}) by $e^{M\sigma z_0}$ to find
\begin{equation}
\frac{d }{dz_0}\,\left[ e^{M\sigma z_{0}} \phi_s \right] = 
M^{3}\sigma \left[ \ln z_{0} - \ln (1-z_{0}) \right]\,e^{M\sigma z_{0}},
\label{intermediate_T_1}
\end{equation}
which allows the integration to be straightforwardly carried out. One finds
\begin{eqnarray}
& & \hspace{-1.5cm} T_{s}(z_0) =  c_{1} e^{-M\sigma z_{0}} + c_2 \nonumber \\  
& & \hspace{-1.5cm} + M^{3}\sigma \int^{z_0}_{0} dy\,e^{-M\sigma y}\,
\int^{y}_{0} dx\,e^{M\sigma x}\,\left[ \ln x - \ln (1-x) \right]  ,
\label{T_s_1}
\end{eqnarray}
where $T_s$ is such that $dT_s/dz_0 = \phi_s$ and $c_1$ and $c_2$ are integration constants. Before imposing the boundary conditions, we can simplify the double integral by differentiating the inner integral and integrating by parts. This gives
\begin{eqnarray}
\hspace{-0.5cm} T_{s}(z_0) =  \hspace{-0.8cm} && c_{1} e^{-M\sigma z_{0}} + c_2 \nonumber \\
             &-& M^2 e^{-M\sigma z_0} \int^{z_0}_{0} dx\,e^{M\sigma x}\,\left[ \ln x - \ln (1-x) \right] \nonumber \\
	      &+& M^2 \int^{z_0}_{0} dy\,\left[ \ln y - \ln (1-y) \right].
\label{T_s_2}
\end{eqnarray}
The last term in Eq.~(\ref{T_s_2}) is simply the $s=0$ mean time to fixation, and so applying the boundary conditions one obtains the equations (\ref{T_linear_in_s}) and (\ref{c_2}) given in the main text.

The calculation of $T(z_0)$ when $\bar{A}(z_0)$ is taken to order $s^2$ can be carried out in a similar way, but the results are more complicated and an integration by parts cannot straightforwardly simplify the double integral down to a single integral. The analogous equation to (\ref{phi_eqn}) is 
\begin{equation}
\frac{d\phi}{dz_0} + M\sigma \left( 1 - s \kappa z_0 \right) \phi = - \frac{M^2}{z_{0}(1-z_{0})},
\label{Phi_eqn}
\end{equation}
where $\kappa=k_{2}/k_{1}$ and $\sigma$ is now given by $\sigma = k_{1}s/\sqrt{b_{1}}$. This is just as singular as Eq.~(\ref{phi_eqn}), and so we perform the same manoeuvre and write $\phi =\phi_{0} + \phi_{s}$, choosing $\phi_0$ so that the term $d\phi_0/dz_0$ cancels the right-hand side of Eq.~(\ref{Phi_eqn}). The equation for $\phi_s$ then reads
\begin{equation}
\frac{d\phi_s}{dz_0} + M\sigma \left( 1 - s \kappa z_0 \right) \phi_s = M^{3}\sigma 
\left( 1 - s \kappa z_0 \right)\,\left[ \ln z_{0} - \ln (1-z_{0}) \right].
\label{Phi_eqn_1}
\end{equation}
The right-hand side is now less divergent, and one can proceed as before to multiply this equation by $e^{M\sigma(z_0 - s\kappa z^{2}_{0}/2)}$ and integrate twice. We find
\begin{eqnarray}
& & \hspace{-1.5cm} T(z_{0}) = - M^{2}\left[ z_{0}\ln(z_{0}) + 
(1-z_{0})\ln(1-z_{0}) \right] \nonumber \\
& & \hspace{-1.5cm} + M^{3}\sigma \int_{0}^{z_{0}} 
dy\,e^{-M \sigma y(1 - s \kappa y /2)} \left\{ \int_{0}^{y} dx (1 - s \kappa x)  
\right. \times \nonumber \\
& & \hspace{-1.5cm} \left. e^{M \sigma x(1 - s \kappa x /2)} 
\int_{0}^{y} \left[ \ln x - \ln(1-x) \right]  - c_{3} \right\}\, ,
\end{eqnarray}
where the constant $c_{3}$ is given by 
\begin{eqnarray*}
& & \hspace{-1.3cm} c_{3} =  \left( \int_{0}^{1} 
dy e^{-M \sigma y(1 - s \kappa y /2)}  \right)^{-1}\,\int_{0}^{1} dy\, 
e^{-M \sigma y(1 - s \kappa y /2)} \nonumber \\
& & \hspace{-0.7cm} \times \int_{0}^{y} dx\,(1 - s \kappa x)\,
e^{M \sigma x(1 - s \kappa x /2)} \left[ \ln x - \ln(1-x) \right] \, .
\end{eqnarray*}
 
\end{appendix}

\end{document}